\documentclass[10pt,a4paper,twocolumn]{article}

\usepackage{lineno}
\usepackage{fontspec}
\usepackage{unicode-math}
\usepackage{authblk}
\usepackage{graphicx}
\usepackage[explicit]{titlesec}
\usepackage[labelfont=bf,labelsep=endash,font=footnotesize]{caption}
\usepackage{tabu}
\usepackage{xcolor}
\usepackage{colortbl}
\usepackage[
    backend=biber, 
    natbib=true,
    style=numeric,
    maxnames=50,
    sorting=none
]{biblatex}
\usepackage[hidelinks]{hyperref}
\usepackage[hang]{footmisc}
\usepackage[normalem]{ulem}
\usepackage[top=2.5cm, bottom=2.8cm, left=1.5cm, right=1.5cm]{geometry}
\usepackage{abstract}
\usepackage{balance}
\usepackage{wrapfig}
\usepackage{enumitem}
\usepackage{soul}
\usepackage{subcaption}
\usepackage{makecell}
\usepackage[export]{adjustbox}
\usepackage{cancel}

\makeatletter
\renewcommand\AB@affilsepx{, \protect\Affilfont}
\makeatother

\providecommand{\keywords}[1]{\textbf{Keywords}\ \ \textendash\ \   #1}

\titleformat{\section}{\large\bfseries}{\thesection.}{1em}{\MakeUppercase{#1}}
\titlespacing*{\section}{0pt}{12pt}{6pt}

\titleformat{\subsection}{\large}{\thesubsection}{1em}{#1}
\titlespacing*{\subsection}{0pt}{12pt}{6pt}

\titleformat{\subsubsection}{\large\itshape}{\thesubsubsection}{1em}{#1}
\titlespacing*{\subsubsection}{0pt}{12pt}{6pt}

\newcommand{\ITUurl}[1]{\textcolor{blue}{\urlstyle{same}\url{#1}}}

\newcommand{\ITUpar}{\vspace{8pt}\par}

\renewenvironment{abstract}
               {\list{}{
               \setlength{\rightmargin}{0mm}
               \setlength{\leftmargin}{0mm}
               \vspace{-0.25in}
                \item[\textit{\textbf{\hspace{22pt}Abstract  }}  \textendash]\relax}}
               {\endlist}

\usepackage[none]{hyphenat}

\usepackage[normalem]{ulem}
\newif\ifrev
\revfalse 
\ifrev
\newcommand{\rev}[1]{\textcolor{blue}{#1}}
\else
\newcommand{\rev}[1]{#1}
\fi

\def\starttable{\vspace{6pt}\begin{table}[ht]\center}
\def\startfigure{\vspace{6pt}\begin{figure}[ht]\center}

\makeatletter
\def\tagform@#1{\maketag@@@{\ignorespaces#1\unskip\@@italiccorr}}
\makeatother

\title{\large{\textbf{\uppercase{Open-Source Emulation-Based Test Environment to Settle O-RAN-Compliant Trials}}}}
\author{Ramon Fontes, Allan Martins, Vicente Sousa, Kaio Dantas, Lucas Medeiros, Pedro Alves, Marcelo Fernandes, Iago Rego, Eduardo Aranha, Vinícius Filho, Mateus Goldbarg, Wysterlanya Barros, Roger Immich, and Augusto V. Neto\\\
\\
\textit{Federal University of Rio Grande do Norte (UFRN), Natal, Brazil}\\\
\\
ramon.fontes@imd.ufrn.br, allan@dca.ufrn.br, vicente.sousa@ufrn.br, kaio.henrique@ifrn.edu.br, \\ 
\{lucas.medeiros.114, pedro.alves.700\}@ufrn.edu.br,~mfernandes@dca.ufrn.br, \\ 
iago.diogenes.072@ufrn.edu.br, eduardo.aranha@ufrn.br, \{vjmtbf, mateus.goldbarg, wysterlanya.barros.016\}@ufrn.edu.br, roger@imd.ufrn.br, augusto@dimap.ufrn.br
}

\begin{document}


\twocolumn[

\begin{@twocolumnfalse}
\date{}\maketitle

\begin{abstract}
\textit{Experimental tools are \rev{a} key \rev{factor in} both academic and industrial research communities to \rev{create} design evaluations \rev{of} new networking technologies \rev{that involve} \rev{troubleshooting} or \rev{changing} the planning of deployed networks. Physical Software-Defined Radio (SDR) experimental platforms enable a design solution for the quick prototyping of wireless communication systems. However, SDR-based experimental platforms \rev{incur high costs}, which \rev{leads} to scalability limitations \rev{in} the experimental settings. Having \rev{said} this, network simulators, emulators, and new testbeds have \rev{attracted} increasing attention. Emulation-based research prototyping \rev{can be distinguished} from real communication networks and SDR-based platforms by allowing a tradeoff between cost and flexibility. This paper \rev{examines} the Mininet-RAN emulation tool, which, \rev{as well as} Radio Access Network (RAN) modeling, provides a way to test Open RAN Intelligent Controller (RIC) services without the need to deploy an entire RAN infrastructure. The Mininet-RAN creates virtual network elements, such as hosts, L2/L3 devices, controllers, and links, by combining \rev{some} of the best emulator features, hardware testbeds, and simulators. By running the \rev{current} code of standard \rev{practice} Unix/Linux network applications and network stack, the Mininet-RAN enables real\rev{-world} network data traffic patterns \rev{to be} delivered to the RIC, regarding the most \rev{significant} aspect of the dynamic generation of wireless system’s KPIs. We provide the basic code of Mininet-RAN for the first two O-RAN Alliance-defined use cases involving V2X and UAV. The xApps are \rev{being} implemented in O-RAN SC near-RT RIC, with Mininet-RAN \rev{which provides}  a closed-loop validation environment.}
\end{abstract}

\ITUpar

\ITUpar
\ITUpar

\end{@twocolumnfalse}
]

\section{Introduction} 
\label{sec:introduction}

\ITUpar Fifth-generation (5G) mobile networks, and \rev{more advanced} technologies, have established disruptive \rev{innovations} to the wireless networking landscape, including vertical services from low-energy/high-dense IoT to delay/reliability-sensitive applications~\cite{necos-commag}. \rev{Apart  from this}, IEEE-based systems are evolving toward high-spectral efficient 802.11be (Wi-Fi 7), \rev{resulting in } a significant \rev{enhancement of} performance \rev{of} the previous 802.11ax (Wi-Fi 6), \rev{as well as } in terms of latency.

\ITUpar As a revolutionary and innovative \rev{measure}, a set of important telecom players created the O-RAN Alliance to promote more democratic and permission-less telecom systems connected via open interfaces and multivendor interoperable components. For many vendors, operators, \rev{policy makers}, and telecom ecosystem stakeholders, the O-RAN movement is crucial to the future of the 3GPP umbrella system, including 4G, 5G, and Wi-Fi~\cite{whitepaperparallel2022}. \rev{These kinds of O-RAN} stakeholders \rev{welcome} the potential of this openness initiative to improve competition, network flexibility, cost \rev{effectiveness}, and features through a centralized Radio Access Network (RAN) Intelligent Controller (RIC) and data-driven closed-loop control~\cite{O-RAN-NFV-SDN-22}.

\ITUpar \rev{The conceptualization} and adoption of new technologies constantly need new experimentation platforms \rev{to ensure trials can be carried out with enhanced and agile deployments}. \rev{However,} before researchers can move forward with full-scale projects \rev{to} drive \rev{their} commercial deployment,  \rev{the project} can be assessed upon pilot studies supported by experimental platforms, such as network simulation and emulation tools~\cite{O-RAN-SBRT-22}. \rev{The reason for this is that the 5G/O-RAN technological system is not an off-the-shelf commercial product and involves the investigation of fundamental aspects (e.g., feasibility, cost, and potential problems).}

\ITUpar Researchers have built testbeds that help to \rev{better understand and demonstrate} the operational capabilities of targeting technologies. However, developing and maintaining a full-scale testbed is expensive \rev{and} time-consuming. The management of \rev{these kinds of} testbed settings is complex \rev{because of} the large physical area required to set up any \rev{useful} multihop topology. \rev{What is worse}, it is \rev{also} difficult to reconfigure such a testbed and even replicate it in other projects.

\ITUpar \rev{For this reason}, both simulation and emulation tools have been \rev{welcomed} by researchers in the academic community and in part of the \rev{industrial world}. \rev{While the exact quantification of each characteristic and the degree of realism ultimately depend on the accuracy of the model implemented in each specific tool among other platform aspects that may affect each feature, Table~\ref{ranking} aims to illustrate the main strengths and shortcomings
typically common to each type of experimentation approach as a first guide to choose the
best type tool for a given set of research goals and constraints.}

\rev{
\begin{itemize}[noitemsep]
    \item \textit{Total cost}: evaluates the cost of experiments, \rev{especially} those related to hardware
and software costs.
    \item \textit{Overall fidelity}: capacity of transferability of the results, accuracy, conclusions,
and the study environment into the real world.
    \item \textit{Replay real traces}: capacity of replaying observed network behavior, such as,
signal strength, throughput, latency, mobility, etc.
    \item  \textit{Real applications}: ability to run real applications without modifying the source
code and with no additional effort by the user hand-side.
    \item \textit{Traffic realism}: assesses the capability of generating, receiving, and processing real
traffic.
    \item \textit{Timing realism}: analyze whether the timing behavior of the system is close to
the behavior of deployed hardware.
    \item \textit{Scalability}: assesses the feasibility of large-scale experiments with respect to the
number of nodes, the experiment duration, and the number of network connections
during the experiment.
    \item \textit{Maintainability}: describes the ability to maintain the evaluation environment. In
other words, \rev{the effort} necessary to keep the system runnable.
    \item \textit{Flexibility}: describes the freedom in creating different experiment scenarios (e.g.
network topology, number of nodes, etc).
    \item \textit{Replication}: how straightforward the repetition of a given experiment in a specific
study environment is.
    \item \textit{Isolation}: assesses the degree \rev{to whose} links, queues, and switches a network \rev{behaves}.
\end{itemize}
}

\rev{The} most appropriate approach \rev{that can be adopted} in experimentally-driven research endeavors is to seek to evaluate the functionality and performance of a network, \rev{and this} always \rev{involves} a tradeoff~\cite{Mininet-NFV}. On \rev{the} one hand, a simulator enables fully controlled application testing-purpose software environments to be set up quickly and easily, with the hardware \rev{layer}. On the other hand, an emulator \rev{moves} further towards close-to-real scenarios by enabling both real-world software and hardware technologies \rev{to be set up} in an integrated environment. 

\begin{table}[!t]
\centering
\caption{Ranking of simulators, emulators and testbeds (adapted from \cite{Zimmermann:2006:AHM:1160987.1161004}).}
\label{ranking}{
\begin{tabular}{*{4}{|c}|}
\hline
\multicolumn{1}{|c|}{\textbf{Characteristic}} & \multicolumn{1}{c|}{\textbf{Sims}} & \multicolumn{1}{c|}{\textbf{Emuls}} & \multicolumn{1}{c|}{\textbf{Testbeds}}   \\ 
\hline
\cline{2-4} Total Cost & $\bullet$$\circ$$\circ$ & $\bullet$$\circ$$\circ$ & $\bullet$$\bullet$$\bullet$ \\
\cline{2-4} Overall Fidelity & $\bullet$$\circ$$\circ$ & $\bullet$$\bullet$$\circ$ & $\bullet$$\bullet$$\bullet$ \\
\cline{2-4} Replay Real Traces & $\bullet$$\bullet$$\circ$ & $\bullet$$\bullet$$\circ$ & $\bullet$$\bullet$$\bullet$ \\
\cline{2-4} Real Applications  & $\bullet$$\circ$$\circ$ & $\bullet$$\bullet$$\bullet$ & $\bullet$$\bullet$$\bullet$ \\
\cline{2-4} Traffic Realism  & $\bullet$$\circ$$\circ$ & $\bullet$$\bullet$$\bullet$ & $\bullet$$\bullet$$\bullet$ \\
\cline{2-4} Timing Realism & $\bullet$$\bullet$$\bullet$ & $\bullet$$\bullet$$\circ$ & $\bullet$$\bullet$$\bullet$ \\
\cline{2-4} Scalability & $\bullet$$\bullet$$\bullet$ & $\bullet$$\bullet$$\circ$ & $\bullet$$\circ$$\circ$ \\
\cline{2-4} Maintainability & $\bullet$$\bullet$$\bullet$ & $\bullet$$\bullet$$\bullet$ & $\bullet$$\circ$$\circ$ \\
\cline{2-4} Flexibility & $\bullet$$\bullet$$\bullet$ & $\bullet$$\bullet$$\bullet$ & $\bullet$$\circ$$\circ$ \\
\cline{2-4} Replication & $\bullet$$\bullet$$\bullet$ & $\bullet$$\bullet$$\bullet$ & $\bullet$$\circ$$\circ$\\ 
\cline{2-4} Isolation & $\bullet$$\bullet$$\bullet$ & $\bullet$$\bullet$$\circ$ & $\bullet$$\bullet$$\bullet$ \\
\hline
\end{tabular}
}
\end{table}

\ITUpar Even \rev{when} its advantages \rev{are taken into account}, designing an O-RAN-compliant emulator to \rev{perform} RIC experimentation is not a trivial task. The fact is that, up to now, there \rev{has been} a lack of open-source emulation tools that are accessible to the general public. In order to fill this gap, we advance beyond the state of the art by designing a new open-source and easy-to-use emulation framework, which caters to end-to-end O-RAN close-to-real testing at a low cost, along with a moderate learning curve. \rev{This kind of} proposal, \rev{called} Mininet-RAN, \rev{takes precedence} over other tools while an emulation platform, by enabling new O-RAN-compliant features to be implemented, evaluated, and validated atop a \rev{trustworthy} O-RAN RIC environment. The \rev{research} contributions of this paper are as follows:

\begin{itemize}
    \item providing insights \rev{into} existing simulation and emulation solutions used for a O-RAN RIC experimentation;
    \item designing an emulation tool capable \rev{of settling} low-cost evaluation and validation trials of O-RAN RIC scenarios under real network traffic and \rev{wireless} technology patterns;
    \item providing a closed-loop experimental use case scenario atop the O-RAN RIC approach;
    \item making available all the codes, of both Mininet-\rev{RAN} and use case scenarios, in an open-source repository for global-community access.
\end{itemize}

\ITUpar The paper is \rev{structured} as follows. Section~\ref{sec_ORAN} provides an overview of the O-RAN RIC technology; Section~\ref{sec_relatedworks} provides \rev{an} analysis and \rev{ideas} of key related work; Section~\ref{sec_mnWiFi} introduces the Mininet-RAN tool, along with the system\rev{'s} architecture; Section~\ref{sec_case} features Mininet-RAN \rev{and employs} two use cases; Section~\ref{sec_limitations} discusses \rev{the} current limitations \rev{of the research} \rev{and} ongoing work\rev{, and makes suggestions for future studies}. Finally, Section~\ref{sec_conclusions} wraps up the paper with \rev{some} final remarks and \rev{encourages} the reader to follow the developments in the open-source code repository.

\section{Overview of  O-RAN RIC}
\label{sec_ORAN}

\ITUpar \rev{Aligned with 3GPP systems, the RIC is a central component of open RAN. The Open RAN has two main objectives: (i) to provide multivendor interoperability across different hardware and software components in a telecommunication system; and (ii) to create an open ecosystem for developing, deploying, and operating third-party applications (called xApps) to support Radio Resource Management (RRM), higher layer procedure optimization, policy optimization in RAN, and provide guidance, parameters, policies, and Artificial Intelligence/Machine Learning (AI/ML) models to cut costs, improve QoS, and generate new streams of revenue.}

\begin{figure}[!t]
    \centering
    \includegraphics[width=\linewidth]{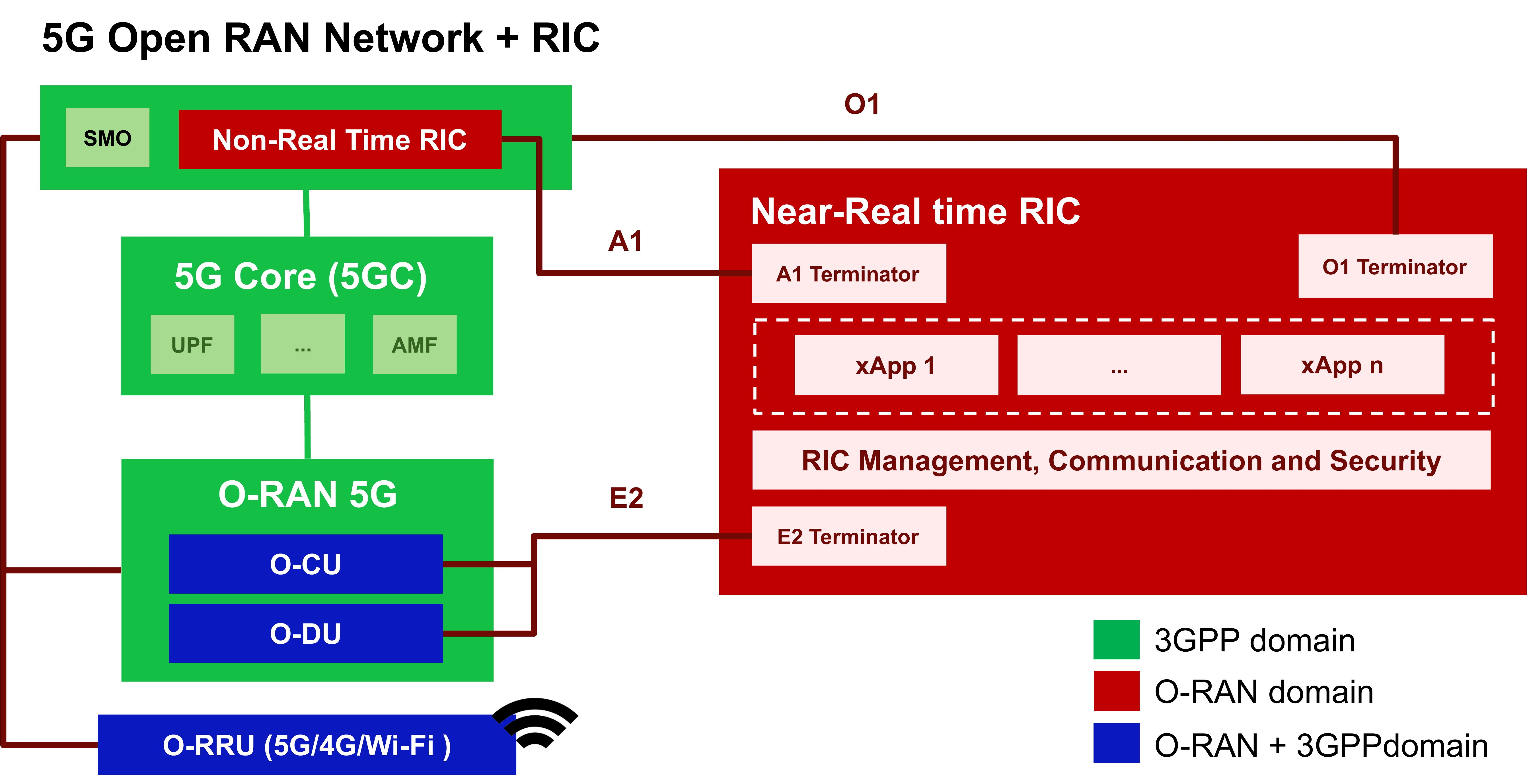}
    \caption{\label{fig_RIC} O-RAN RIC components (adapted from https://www.o-ran.org/).}
\end{figure}

\ITUpar To unlock \rev{limitless} innovation, the O-RAN Alliance, a worldwide community of mobile network operators, vendors, and R\&D institutions, defines a reference architecture \rev{comprising} of the near-Real-Time (near-RT) RIC, non-RT RIC, and A1, E2, and O1 interfaces. Fig.~\ref{fig_RIC} illustrates the O-RAN capable 3GPP systems \rev{that exchange} Key Performance Indicators (KPIs) \rev{with} the RIC \rev{to} evaluate and control 3GPP components.

\ITUpar \rev{These} RAN control strategies can be carried out in a non-RT control loop (operating \rev{on} a scale greater than 1000 ms) and near-RT control loop (performing operations between 10 ms and 1000 ms). The xApps are crucial elements of near-RT RIC. As \rev{they are} a set of microservices, xApps can make intelligent decisions for adapting RAN parameters (e.g., subscriber positioning, handover to a cell, \rev{changes} to different carrier frequencies) to optimize the subscriber experience and network performance. Thus, RIC xApps could request, via \rev{the} E2 interface, KPIs from RAN, then send back control actions across the E2 interface to the RAN.

\ITUpar \rev{Unlike} the traditional RAN \rev{evolutionary} approach, a system architecture with the RIC \rev{is flexible enough} to extend the RRM actions by implementing third-party microservices without changing the RAN implementation regardless of its brand. \rev{Thus, through the O-RAN set of protocols, a proprietary RAN that supports the RIC} has a two-way communication channel through which the RIC can measure and control the RAN transmission functionality. The novelty is that anyone can provide those third-party implementations by xApps and not only the RAN proprietary companies.

\ITUpar With the \rev{supplementary} operation of both near-RT and non-RT RIC loops, complete ML-powered management can be deployed, \rev{which covers} optimization actions from lower to upper layers protocols. They may \rev{range from} short-term MAC-layer procedures (e.g., scheduling) to long-term network decisions like network slice assurance to provide end-to-end connectivity \rev{and} quality of service. While near-RT RIC monitors and rapidly regulates \rev{the} RAN actions based on ML models, the non-RT RIC \rev{ensures} the best-trained ML models and provides suitable policies to guide the \rev{operational performance} of near-RT RIC functions. \rev{More specifically, the non-RT RIC takes advantage of the Service Management and Orchestration (SMO) framework to provide policy-based guidance, \rev{strategies for} model management, and enrichment information to the near-RT RIC function in response to specialized applications called rApps.}

\section{Related Work}
\label{sec_relatedworks}

\ITUpar \rev{This section gives} an overview of the most \rev{significant} footprints in network experimentation platforms, \rev{and highlights} the main \rev{features to obtain}  a better understanding of the tradeoffs when \rev{seeking} to support realistic O-RAN experimentation. It is worth \rev{noting} that a few third-party tools are available for purchase, such as the VIAVI's TeraVM\footnote{http://rb.gy/qahxk6} and the Keysight's P8828S\footnote{http://rb.gy/p56gft}, which are solutions that emulate O-RAN Alliance WG3-standardized E2 node types for setting up RIC test trials \rev{in} different scales. \rev{However, as these} third-party solutions are \rev{very} expensive, they \rev{are beyond} the scope \rev{of this paper}, which focuses on low-cost and open-source endeavors.

Table~\ref{tab:papers} depicts the list of the most relevant 5G and O-RAN-tailored open-source simulator and emulator initiatives we found in the literature. For each tool, we raise aspects ranging from physical layer modeling to End-to-End (E2E) evaluation, along with software licensing, the last code update, and the availability of the O-RAN E2-interface highlights.

\begin{table*}[!t]
    \centering
    \caption{Summary of open-source simulators/emulators for O-RAN end-to-end (E2E) performance evaluation.}
    \label{tab:papers}
    \resizebox{\linewidth}{!}{%
	\begin{tabular}{cccccc}
	\hline 
	\multicolumn{1}{c}{\footnotesize \textbf{\makecell{Experimentation\\Platform}}} &
    \multicolumn{1}{c}{\footnotesize \textbf{\makecell{Complete PHY \\ abstraction (channel \\and error models)}} }&
    \multicolumn{1}{c}{\footnotesize \textbf{\makecell{E2E \\Evaluation}}} &
    \multicolumn{1}{c}{\footnotesize \textbf{\makecell{O-RAN \\Interfacing}}} & 
    \multicolumn{1}{c}{\footnotesize \textbf{\makecell{Open \\ Source}}} & {\footnotesize 
	\textbf{\makecell{Last \\ Update}}} \\ \hline
\rowcolor[HTML]{EFEFEF}	\footnotesize \begin{tabular}[c]{@{}l@{}} OAI L1 RF \\ Simulator~\cite{OAI_RAN_site} \end{tabular} & \footnotesize \begin{tabular}[c]{@{}l@{}} only RF (radio channel simulator) \end{tabular}
   & \footnotesize \begin{tabular}[c]{@{}c@{}} yes, with OAI infra \end{tabular} & \footnotesize no & \footnotesize  \begin{tabular}[c]{@{}c@{}} OAI Public \\ License V1.1 \end{tabular}  & \footnotesize \makecell{Aug, 2021\\}
    \\

    \footnotesize \makecell{OAI L2 nFAPI \\ Simulator~\cite{OAI_RAN_site}} & \footnotesize \makecell{using OAI L1 RF simulation}
   & \footnotesize \makecell{yes, with  OAI Infra} 
 & \footnotesize no & \footnotesize  \makecell{OAI Public \\ License V1.1} & \footnotesize \makecell{Aug, 2022\\}
    \\
  \rowcolor[HTML]{EFEFEF}  {\footnotesize simu5G~\cite{simu5G_site}}  & \footnotesize yes   & \footnotesize \begin{tabular}[c]{@{}c@{}}user plane only \\ control plane \\ not modeled ~\cite{simu5G_paper} \end{tabular}  & \footnotesize no & \footnotesize LGPL  & \footnotesize Oct, 2022 \\
    {\footnotesize UERANSIM~\cite{UERANSIM_RANSim_repo}}  & \footnotesize \begin{tabular}[c]{@{}c@{}}no radio protocols \\ below the RRC layer \end{tabular}  & \footnotesize no & \footnotesize no & \footnotesize GPL-3.0 & \footnotesize Oct, 2022 \\

\rowcolor[HTML]{EFEFEF} \footnotesize 5G Lena ns-3~\cite{lena_ns3_site}            & \footnotesize \begin{tabular}[c]{@{}c@{}} yes, without handover and \\ mobility (NSA only)~\cite{lena_ns3_paper} \end{tabular}
  & \footnotesize \begin{tabular}[c]{@{}c@{}} yes, simulation \\(no real time) \end{tabular}
 & \footnotesize no & \footnotesize GNU GPLv2 & \footnotesize Nov, 2022\\    

{\footnotesize FikoRE~\cite{GonzalezD2022}}  & \footnotesize \makecell{focus on application-level \\ experimentation and prototyping
}   & \footnotesize \makecell{yes (VR/AR devices \\in real-time)} & \footnotesize no & \footnotesize BSD-3-Clause-Clear license  & \footnotesize Feb, 2023\\

\rowcolor[HTML]{EFEFEF}   \footnotesize \begin{tabular}[c]{@{}c@{}} O-RAN SC \\ sim-e2-interface~\cite{oran_e2_site} \end{tabular}  & \footnotesize \begin{tabular}[c]{@{}l@{}} based on VIAVE’s \\simulator dataset~\cite{VIAVE_5G_Sol} \end{tabular}   & \footnotesize \begin{tabular}[c]{@{}l@{}} yes, with O-RAN \\ SC 5G core \end{tabular} & \footnotesize yes & \footnotesize Apache 2.0 & \footnotesize May, 2022 \\

  {\footnotesize \begin{tabular}[c]{@{}c@{}} SD-RAN \\ RAN Simulator~\cite{sdran_RANSim_site} \end{tabular} }  & \footnotesize \begin{tabular}[c]{@{}c@{}} only an E2 agent  using E2AP \end{tabular}   & \footnotesize \begin{tabular}[c]{@{}l@{}} yes, with USRP \end{tabular}  & \footnotesize yes & \footnotesize \begin{tabular}[c]{@{}c@{}} ONF Member-Only  \end{tabular}  & \footnotesize Sep, 2022\\   

\rowcolor[HTML]{EFEFEF}    \footnotesize mmWave ns-3~\cite{mmwave_ns3_paper}            & \footnotesize \begin{tabular}[c]{@{}c@{}} yes (NSA only)  \end{tabular}
  & \footnotesize \begin{tabular}[c]{@{}c@{}} yes, simulation \\(no real time)  \end{tabular}
 & \footnotesize yes & \footnotesize GNU GPLv2  & \footnotesize Oct, 2022\\
    {\footnotesize Mininet-RAN}  & \footnotesize yes
   & \footnotesize yes & \footnotesize yes & \footnotesize GNU GPLv2  & \footnotesize \makecell{Feb, 2023}\\
    \hline
\end{tabular}
}
\end{table*}

\ITUpar The literature reveals \rev{a few} footprints of open-source tools capable of emulating the O-RAN E2-interface for E2E evaluation. Three projects provide both functional software and \rev{an} emulator of O-RAN-ready RAN: O-RAN SC (from O-RAN Alliance), SD-RAN (from Open \rev{Networking} Foundation), and Open Air Interface (EURECOM).

\ITUpar Furthermore, the sim-e2-interface O-RAN SC denotes an E2-interface agent, which only \rev{seeks} to test the E2AP protocol message exchange. Two improvement initiatives to feed RAN-related KPIs into the sim-e2-interface \rev{are ongoing}. The first is a collaboration with VIAVE company (\rev{alternative to} proprietary \rev{software}), while the interface \rev{for} the ns-3 simulator is the low-cost alternative \rev{that involves having} a KPI-based closed-loop with the O-RAN SC emulator.

\ITUpar The SD-RAN provides another option \rev{which is to have} an E2 agent for E2AP, but an E2E option is only possible by \rev{means of} OAI tools. Although \rev{it includes} the SDR-based code of 4G and 5G RANs, the OAI emulates a gNB with an nFAPI emulator. As introduced in~\cite{nFAPI_repo}, the nFAPI defines a network protocol that is used to connect a Physical Network Function (PNF) running \rev{a} PHY layer (Layer 1) to a Virtual Network Function (VNF) running MAC (Layer 2) and above.

\ITUpar On the basis of the information in this section, none of the related solutions described above can cater \rev{for highly-accurate} wraparound test settings for an O-RAN-compliant system workflow. This \rev{lack of} solution \rev{is a motivation factor in} our task in \rev{designing} \rev{a low-cost} and open-source tool to \rev{create} a controllable testbed that \rev{can help}  to exercise and evaluate O-RAN-compliant features quickly, \rev{as well as} seeking to simplify the \rev{task of validating the} new innovations before \rev{they are introduced to the} market.

\rev{Thus, this paper provides an O-RAN compatible experimentation platform \rev{which includes} a complete wireless PHY abstraction (e.g., RF, modulation, channel coding, and antenna aspects), several MAC layer features (e.g., link adaptation, error detection, multiple access), as well as real-time end-to-end service delivery (e.g., video streaming).}

\section{Mininet-RAN Design and Workflow}
\label{sec_mnWiFi}

\begin{figure}[!htb]
    \centering
    \includegraphics[width=\linewidth]{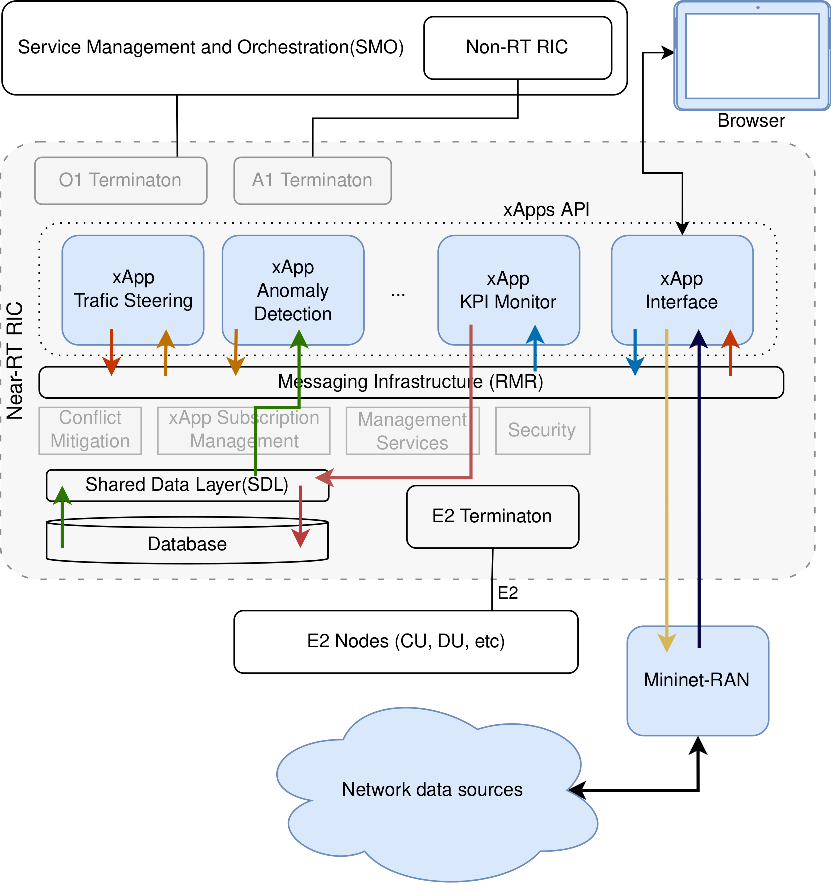}
    \caption{\label{fig_RIC_arch} Proposed architecture for the testbed. }
\end{figure}

\ITUpar This section provides a general description of the proposed solution, in terms of architecture, functionalities, and experimental testbed capabilities. As explained in the previous sections, the O-RAN RIC can run xApps that act as microservices with very specific tasks. In order to test an application \rev{comprising} a set of xApps, it is \rev{necessary} to have \rev{a means of providing} network information (e.g., KPIs) and being able to react with actions (e.g., performing handover) and, more importantly, ensuring that those actions can be reflected in the new measurements. This closed-loop scenario is not trivial and needs some simulation/emulation in real time. For the \rev{purposes} of this work, we propose an architecture together with a particular use case that fulfills this requirement. Fig.~\ref{fig_RIC_arch} shows the complete O-RAN RIC \rev{that is} running with Mininet-RAN.

\ITUpar Fig.~\ref{fig_RIC_arch} shows the proposed near-RT RIC architecture. 
The gray area shows the near-RT RIC connected to both the Service Management and Orchestration (SMO), \rev{which} includes the non-RT RIC, and the E2-related \rev{blocks}. The near-RT RIC includes all xApps and the remaining infrastructure is shown only as a reference. From that infrastructure, the two main parts are the Messaging Infrastructure~(RMR) and Shared Data Layer~(SDL). The proposed architecture comprises four xApps that run on the near-RT RIC and two external applications. 

\ITUpar The xApp highlighted in blue interfaces with the following external elements: (i) Mininet-RAN to collect RAN-related information; and (ii) the dashboard to display data to a browser using a Javascript web app.

\ITUpar As stated before, each xApp runs as a specialized microservice. Their goals and lifecycles are \rev{outlined} as follows:

\begin{itemize}
    \item \textbf{xApp KPI Monitor}:
    \begin{itemize}
        \item goal: 
        to constantly request KPIs from the xApp Interface and update the SDL database with them.
        \item lifecycle: 
        
            \begin{itemize}
                \item Request KPIs via RMR from the xApp Interface;
                \item serialize KPIs and store them in the KEY-VALUE pair in the SDL;
                \item store the KPI data on SDL;
                \item repeat.
            \end{itemize}
        
    \end{itemize}
\end{itemize}

\begin{itemize}
    \item \textbf{xApp Anomaly Detection}:
    \begin{itemize}
        \item goal: 
        to evaluate the KPIs in the SDL based on the trained ML algorithm. The result of this evaluation should \rev{be reflected} in some action 
        to be sent via RMR. 

        \item lifecycle: 
            \begin{itemize}
                \item Read KPIs from the SDL;
                \item process the KPI, generating an action according to the output of the ML algorithm;
                \item send this action via the RMR to whoever is subscribed (normally the xApp Traffic Steering).
            \end{itemize}
    \end{itemize}
\end{itemize}

\begin{itemize}
    \item \textbf{xApp Traffic Steering} 
    \begin{itemize}
        \item goal: 
        to listen for actions that need to be taken and perform them if necessary via xApp Interface.

        \item lifecycle: 
            \begin{itemize}
                \item Listen to the RMR for actions to be performed (normally from the xApp Anomaly Detection;
                \item \rev{check} or filter the action and decide its \rev{significance on the basis of an } algorithm;
                \item send the action to the xApp Interface via the RMR.
            \end{itemize}
    \end{itemize}
\end{itemize}

\begin{itemize}
    \item \textbf{xApp Interface} 
    \begin{itemize}
        \item goal: 
        to serve as an interface with the external software. It communicates with the Mininet-RAN to either ask for KPIs or it \rev{carries out} actions on the emulator. It also \rev{supplies} the external dashboard with network and RIC data.       
        \item lifecycle: 
            \begin{itemize}
                \item Listen for RMR calls to access Mininet-RAN KPI endpoints when requested;
                \item listen for RMR calls to access Mininet-RAN actions endpoints when requested.
            \end{itemize}
    \end{itemize}
\end{itemize}

\subsection{Architecture and implementation}

\rev{Being a network emulator, Mininet-RAN is able to run with virtually all (if not all) tools, programming languages, and other resources supported by Linux systems. In our implementation, network KPIs are collected with the support of Scapy, a packet manipulation library written in Python, which listens for predefined KPIs by the user.
} The communication between Mininet-RAN and the near-RT RIC is performed by REST API. The API has basically two main endpoints: \texttt{/api/v1/getKPI} and \texttt{/api/v1/performHandover}. The former is performed via \texttt{GET} (with no parameters), and it returns the current KPI \rev{measurement}, whereas the latter is performed \rev{through the} \texttt{POST} method. The same approach was adopted to communicate near-RT RIC and the external dashboard.

\subsection{Creating a network}

\ITUpar The network topology script is one of the most important artifacts of Mininet-RAN. To start Mininet-RAN, a Python script is required to bring up the desired network topology. \rev{As a result}, nodes may be created \rev{and}, customized, \rev{while its}  services \rev{are} instantiated. Currently, Mininet-RAN supports wireless technologies such as IEEE 802.11, 
particularly IEEE 802.11p, as well as IEEE 802.15.4. We have also started a prototype with the wwan\_hwsim module\footnote{\url{https://github.com/torvalds/linux/blob/master/drivers/net/wwan/wwan_hwsim.c}}, a Linux Kernel module that we plan to \rev{provide and expect} that it will allow \rev{extensive} experiments \rev{for} LTE and 5G NR settings.

\subsection{Network customization and user interaction}

\ITUpar Mininet-RAN supports all the commands related to Linux system wireless tools such as \texttt{iw}, \texttt{iwpan}, and others, some of them well-known \rev{by} network emulators such as Mininet~\cite{lantz2010network} and Mininet-WiFi~\cite{fontes2015mininet}. For example, the user may use \texttt{node1 iw dev node1-wlan0 scan} to scan for available base stations (BSs), and \texttt{node1 iw dev node1-wlan0 connect ssid} to connect to a selected base station 
with the same SSID. \rev{Since it is} a runtime emulator Mininet-RAN allows users to add new network \rev{features and elements to ensure} a more versatile experiment. To verify the connectivity between virtual devices such as nodes, the user can type the CLI
command: \texttt{node1 ping node2}. Common Linux commands can be executed by the user at runtime, for instance, to \rev{check} the available bandwidth between two nodes (node1 and node2): \texttt{node1 iperf -s \& node2 iperf -c 10.0.0.1}. \rev{As we will see from the next section, Mininet-RAN's versatility makes it possible for this emulator to interoperate with well-known simulators in the community, such as SUMO and CoppeliaSim, bringing more realism to the development of case studies.}

\section{Case Studies}
\label{sec_case}

\ITUpar This section features Mininet-RAN \rev{and employs} two use cases defined by the O-RAN Alliance: (i) the context-based dynamic handover management for V2X; (ii) and the flight path-based dynamic UAV resource allocation. In the context of 5G, we expect Mininet-RAN to \rev{offer more} significant and complementary advantages \rev{than} simulation or testbed-based experimental approaches. \rev{The source code repository including artifacts as well as reproducible results is available at https://github.com/mininet-ran/mininet-ran}.

\subsection{Use case 1: Context-based dynamic handover management for V2X}

\ITUpar 5G NR V2X has been introduced as a standard by 3GPP in Release 16 with advanced functionalities on top of the 5G NR air interface to support connected and \rev{autonomous} driving use cases with stringent requirements. It promises numerous benefits such as increased road safety \rev{and}, reduced emissions, \rev{as well as} saving time by orchestrating the traffic and assisting individual user decisions based on real-time information on the road and traffic conditions, driver intentions, and other \rev{factors}~\cite{boutiba2022nrflex}.

\ITUpar This case study explores context-based dynamic handover management for a V2X scenario. This problem becomes exceptionally challenging when vehicles \rev{experience} frequent handovers. \rev{Owing} to their high speed and the heterogeneous nature of the wireless environment, vehicles might be handed over frequently or in \rev{a} suboptimal ways, which may cause handover anomalies such as short stay, \rev{the} ping-pong effect, and poor radio conditions of the remote cell.

\begin{figure}[!htb]
    \centering
    \includegraphics[width=\linewidth]{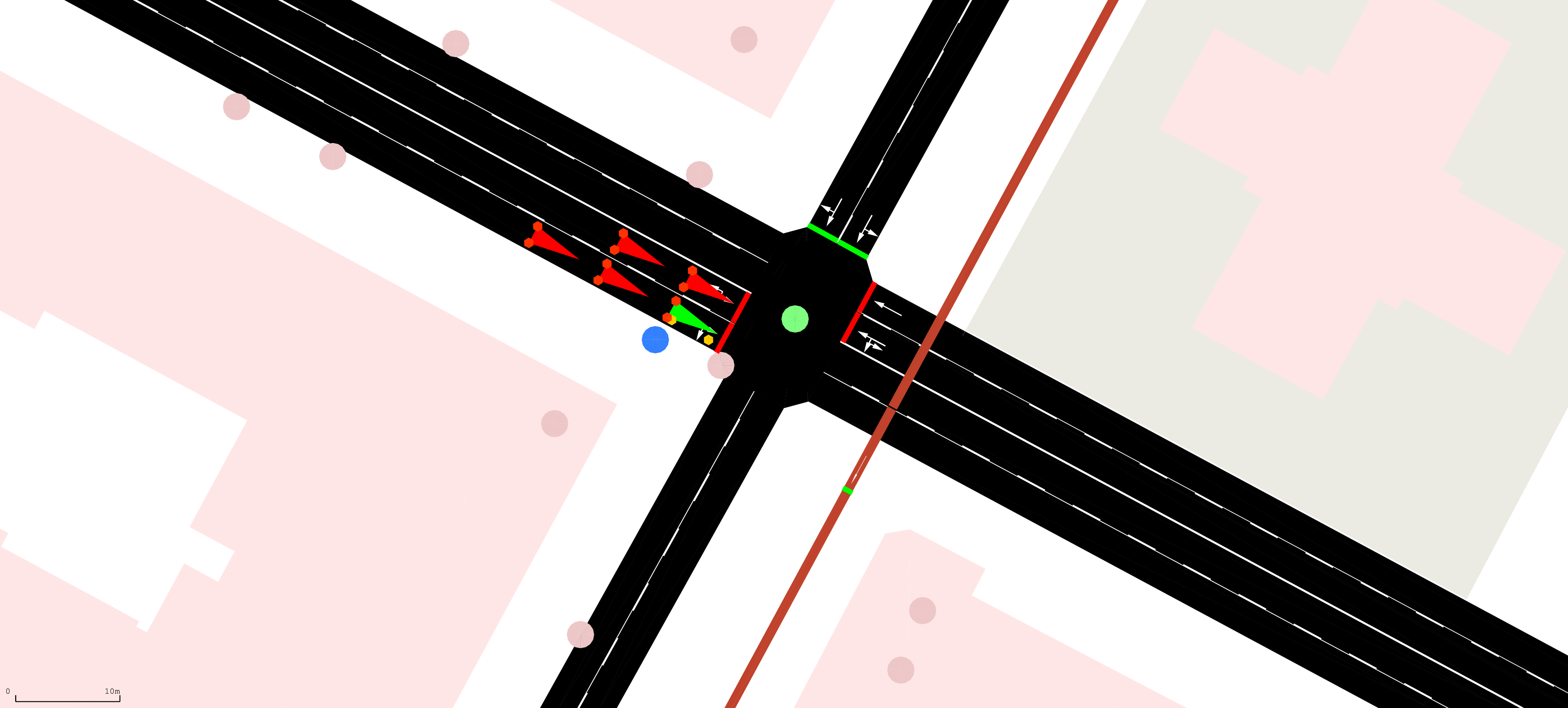}
    \caption{Vehicular Simulation on SUMO.}
    \label{fig:sumo}
\end{figure}

\ITUpar This case study aims to show how versatile Mininet-RAN can be through its integration with O-RAN's near-RT RIC. We \rev{leveraged} Mininet-RAN by introducing a communication path to the near-RT RIC, to generate and share a list of KPIs to drive ML-supported proactive smart handover decisions. More specifically, we use the Open Source Simulation of Urban Mobility (SUMO)~\cite{behrisch2011sumo}, as illustrated in Fig.~\ref{fig:sumo}, to simulate the mobility of vehicles and components of a vehicular network, such as traffic lights and vehicle speed. 

\ITUpar The context-based dynamic handover management for V2X proposed here works as follows:

\begin{itemize}
    \item While vehicles \rev{move around, BSs continuously transmit messages to the near-RT RIC}. The messages contain the Received Signal Strength Indication (RSSI) \rev{picked up} by the base stations;
    \item the near-RT RIC receives and processes the messages sent by the BSs;
    \item the near-RT RIC sends control messages to the base stations \rev{which instruct} vehicles when they have to roam to a BS that provides better signal quality.
\end{itemize}

\ITUpar We now showcase the first set of results and discussions, especially \rev{those} related to ML methods applied to this use case, followed by the \rev{pattern of} behavior of the monitored vehicle using a dashboard. 

\subsubsection{Machine learning of the use case 1}

\ITUpar The dataset used to train the ML in this use case has 900 samples, each consisting of the Signal-to-Noise Ratio (SNR) values of three neighbor BSs (\texttt{BS1}, \texttt{BS2}, \texttt{BS3}), the current connected BS, and a timestamp. The dataset is \rev{divided} into four columns: \texttt{SNR1}, with values referring to the SNR of \texttt{BS1}; \texttt{SNR2}, with values referring to the SNR of \texttt{BS2}; \texttt{SNR3}, with values referring to the SNR of \texttt{BS3}; and \texttt{Current\_BS}, with the indication of the current connected BS. \rev{Since} the goal is to predict the BS switching, the output of each sample should refer to the \texttt{Current\_BS} of the following sample. This output is encoded using the One Hot Encoding technique. It is worth mentioning that the purpose of this use case is to demonstrate the versatility of Mininet-RAN for testing an ML-based near-RT RIC without presenting a deep performance study about the proposed handover.

\ITUpar Our ML approach is a Multilayer Perceptron (MLP) with four inputs (\texttt{SNR1}, \texttt{SNR2}, \texttt{SNR3}, and \texttt{Current\_BS}) and three outputs (the probability of changing the connection for \texttt{BS1}, \texttt{BS2}, and \texttt{BS3}). The preprocessed dataset is split for training and testing the MLP model, with sets containing 70\% and 30\% of the data, respectively.

\ITUpar The MLP architecture is defined \rev{by} the GridSearchCV function, which \rev{employs} hyperparameter tuning to determine the optimal values for a given model. In addition, this technique evaluates the model for each combination using the k-fold cross-validation method. We \rev{analyzed the performance of} the MLP architecture with one or two hidden layers containing 10 to 50 neurons in each layer. While the hidden layers adopt Relu as the activation function, the output layer has Softmax and \rev{shows the} switching probability \rev{of each BS} as the output.

\begin{figure}[!htb]
    \centering
    \includegraphics[width=0.45\textwidth, height=0.4\textheight, keepaspectratio=true]{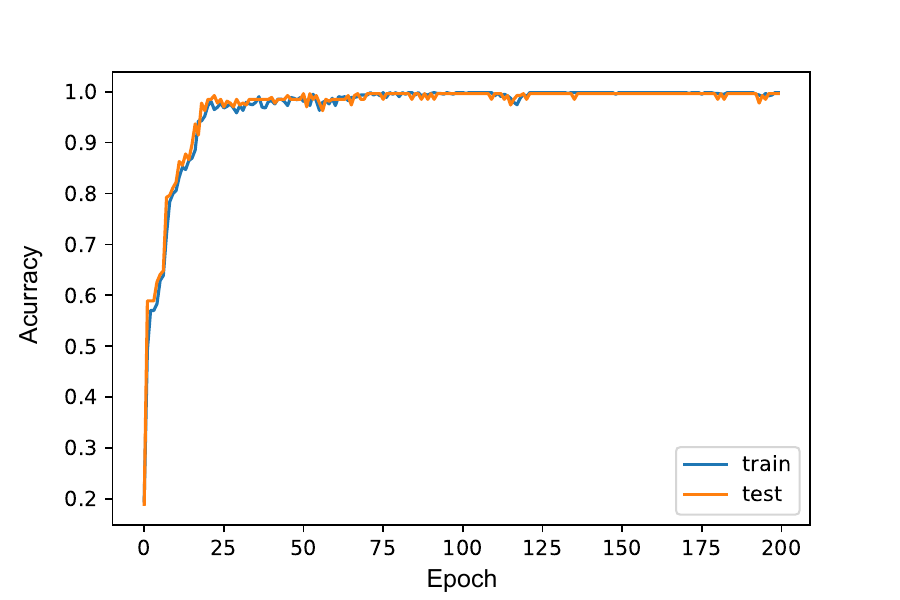}
    \caption{Learning curve.}
    \label{fig:learning_curve}
\end{figure}

\ITUpar The best result from GridSearchCV is for an MLP with two hidden layers of 40 and 20 neurons in the first and second layers, respectively. The k-fold cross-validation with five folds provides 99.84\% of model accuracy as illustrated in the learning curve of Fig.~\ref{fig:learning_curve}. The figure also shows a similar \rev{degree of } accuracy for the test data, \rev{which suggests} that the model is not overfitting and can \rev{be generalized} well \rev{for} new data, i.e., it has learned to \rev{trace} the underlying patterns in the data and can make accurate predictions \rev{about} unseen data.

\begin{figure}[!htb]
\centering
\begin{subfigure}{0.23\textwidth}
    \includegraphics[width=\linewidth]{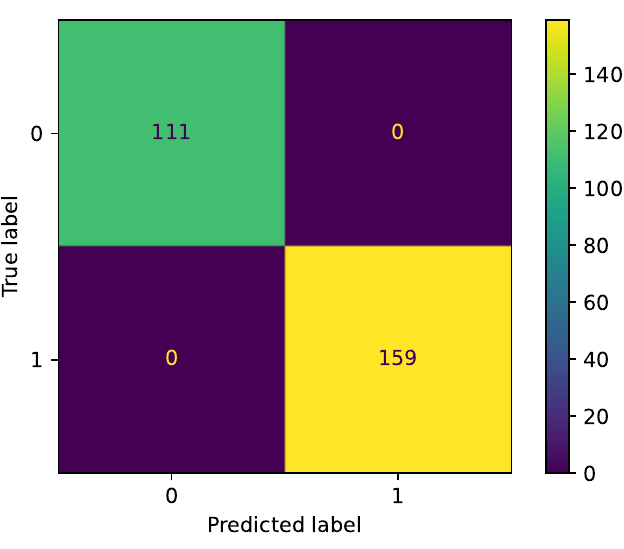}
    \caption{BS1}
    \label{fig:Confusion_Matrix_ENB1}
\end{subfigure}
\hfill
\begin{subfigure}{0.23\textwidth}
    \includegraphics[width=\linewidth]{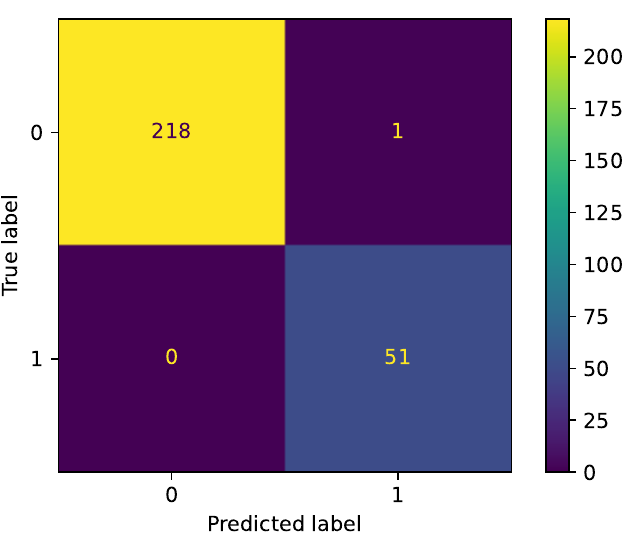}
    \caption{BS2}
    \label{fig:Confusion_Matrix_ENB2}
\end{subfigure}
\hfill
\begin{subfigure}{0.23\textwidth}
    \includegraphics[width=\textwidth]{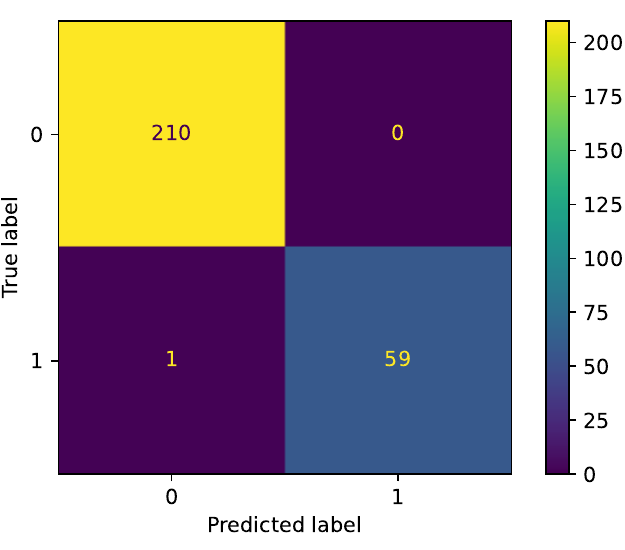}
    \caption{BS3}
    \label{fig:Confusion_Matrix_ENB3}
\end{subfigure}
\caption{Confusion matrices}
\label{fig:confusion_matrix}
\end{figure}

\ITUpar The confusion matrices of the trained MLP are \rev{shown} in Fig.~\ref{fig:confusion_matrix} for the three BSs. According to the results, ML \rev{made} only two mistakes, \rev{which involved} \texttt{BS2} when it was not the best option and \rev{fails} to \rev{involve} \texttt{BS3} when it was the best choice. Figures~\ref{fig:Confusion_Matrix_ENB2} and \ref{fig:Confusion_Matrix_ENB3} \rev{illustrate} those errors, respectively.

\subsubsection{\rev{Results} of use case 1}

\begin{figure}[!htb]
\centering
\begin{subfigure}{0.5\textwidth}
    \includegraphics[width=\textwidth]{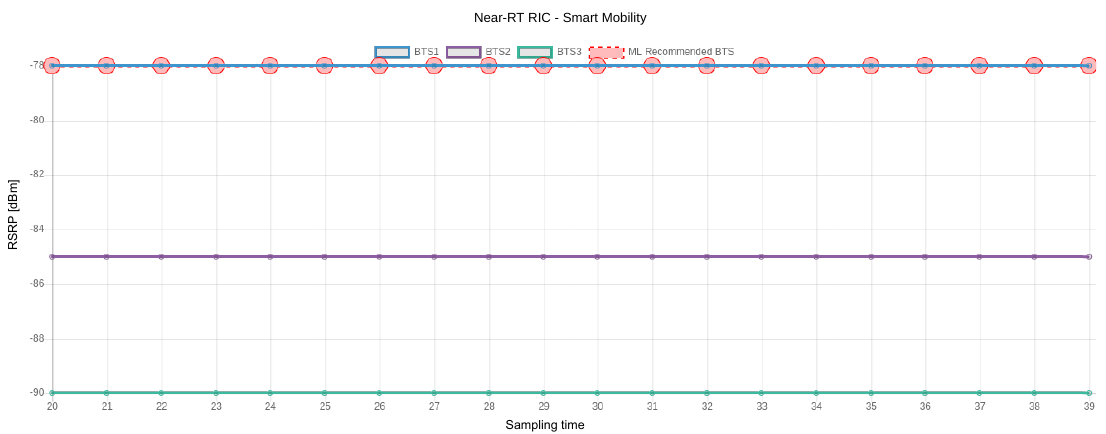}
    \caption{Vehicle 1 connected to BTS1.}
    \label{fig:Dashboard1}
\end{subfigure}
\hfill
\begin{subfigure}{0.5\textwidth}
    \includegraphics[width=\textwidth]{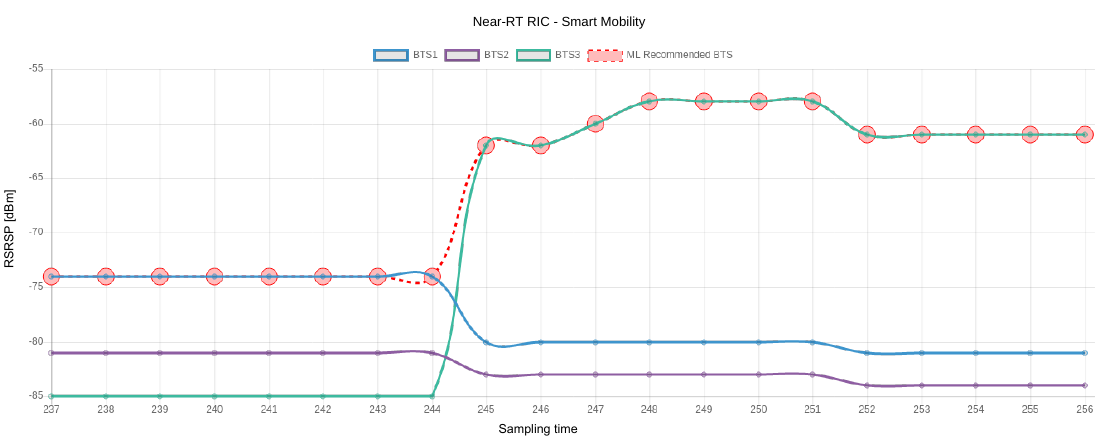}
    \caption{Vehicle 1 connected to BTS3.}
    \label{fig:Dashboard3}
\end{subfigure}
        
\caption{Near-RT RIC dashboard monitoring
information}.
\label{fig:kpi_measurement}
\end{figure}

\ITUpar As illustrated in Fig.~\ref{fig:kpi_measurement}, the vehicle dashboard shows a history of the KPIs' measurement from the three nearest BSs of the monitored vehicle. The three solid lines represent a history for each BS SNR level sensed by the vehicle, while the dotted line with a circle marker \rev{makes clear} which BS \rev{is recommended by } the ML \rev{to best serve the vehicle}.

\begin{figure}[!htb]
\centering
\begin{subfigure}{0.5\textwidth}
    \includegraphics[width=\textwidth]{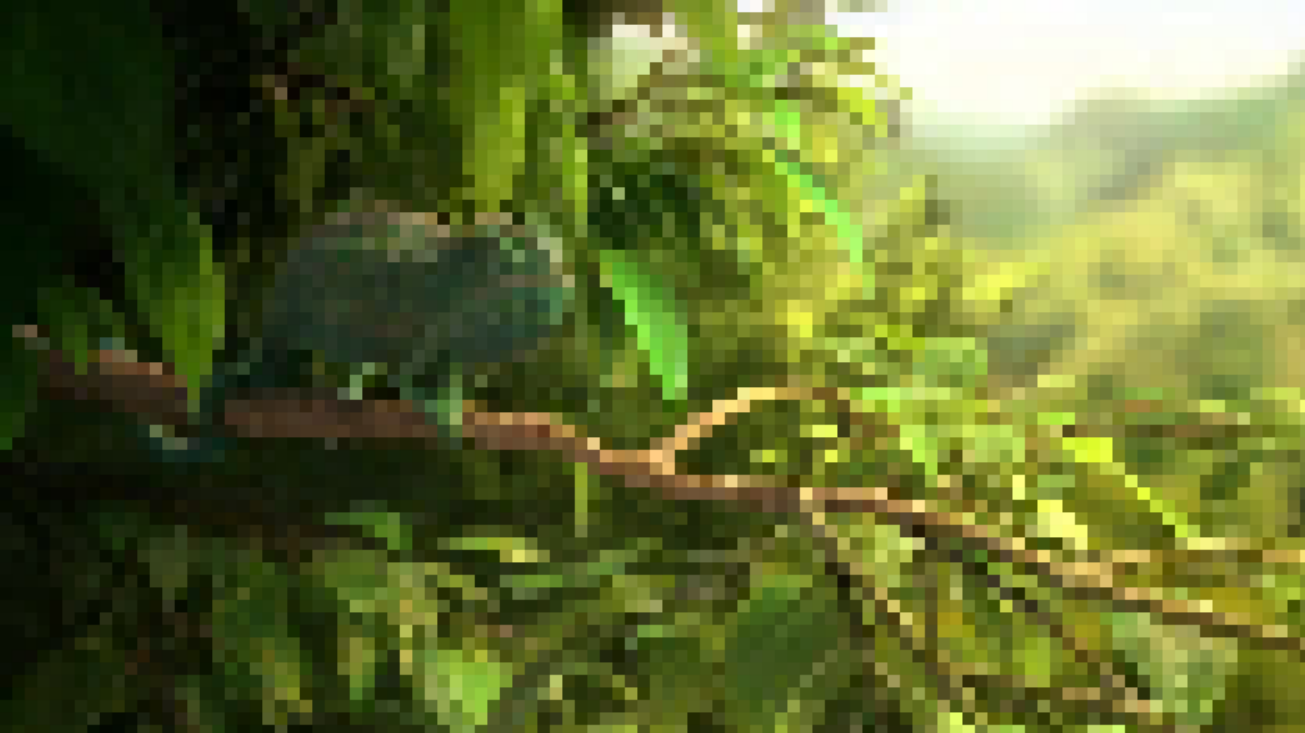}
    \caption{Without near-RT RIC solution enforcement.}
    \label{fig:UC_poor}
\end{subfigure}
\hfill
\begin{subfigure}{0.5\textwidth}
    \includegraphics[width=\textwidth]{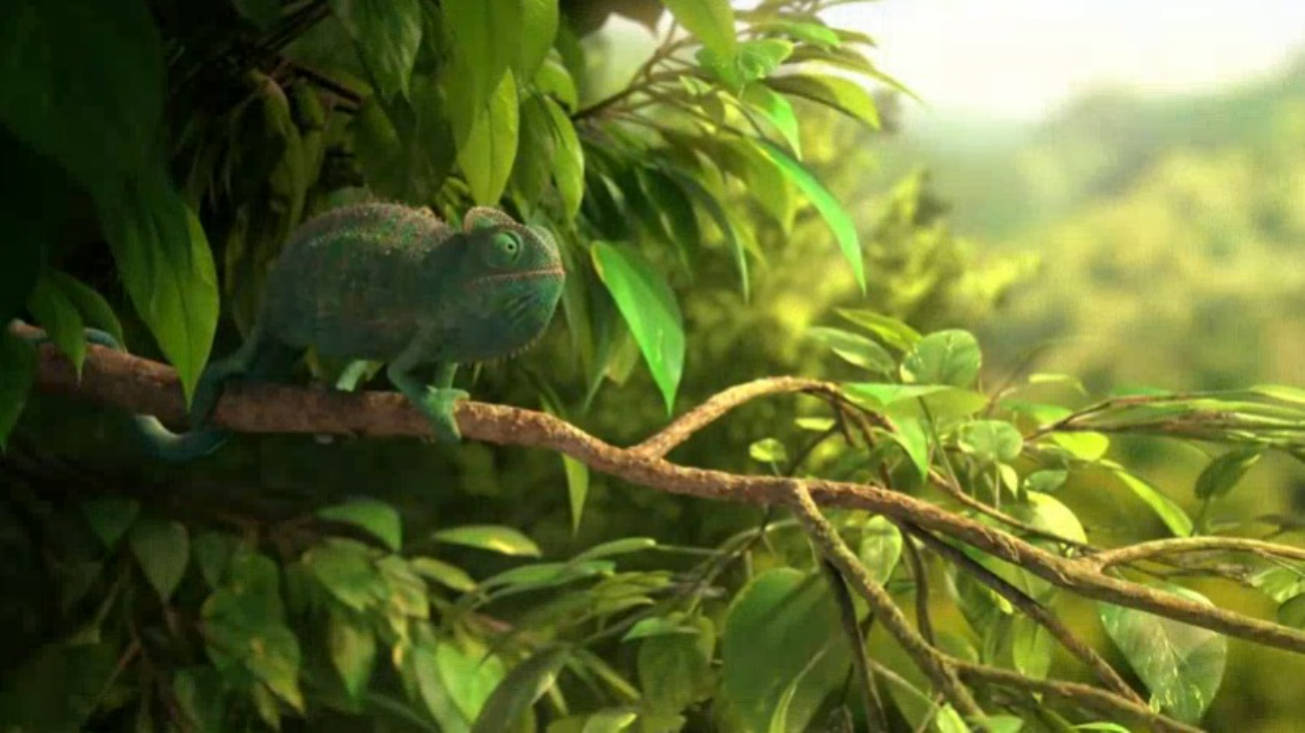}
    \caption{With near-RT RIC solution enforcement.}
    \label{fig:UC1_good}
\end{subfigure}
        
\caption{Quality of videos.}
\label{fig:video-quality}
\end{figure}

\ITUpar To illustrate how the dashboard of this use case works, the emulation started with a vehicle connected to \texttt{BS1}, as shown in Fig.~\ref{fig:Dashboard1}. The vehicle \rev{also obtains} \rev{a stream} of a high-resolution video. As the vehicle moves \rev{along} the route established in the SUMO, the near-RT RIC keeps measuring the available KPIs and feeding the xApp Anomaly detection, resulting in a handover action sent to the vehicle. Fig.~\ref{fig:Dashboard3} illustrates the moment a handover is recommended as the \texttt{BS3} SNR level gets better than the \texttt{BS1} \rev{to which} the vehicle is \rev{connected}. The handover decision is not \rev{made} immediately after a new BS KPI \rev{has improved}, a necessary measure to avoid the ping-pong effect between BSs. \rev{A} difference \rev{can be noticed } between the experiment \rev{both with and } without the near-RT RIC optimization in Fig.~\ref{fig:UC_poor}, as illustrated in Fig.~\ref{fig:UC1_good}.

\subsection{Use Case 2: Flight path-based dynamic UAV resource allocation}

Unmanned Aerial Vehicles (UAVs) are expected to be \rev{an integral part of} the upcoming wireless networks, \rev{by} potentially facilitating wireless broadcasting and supporting high-\rev{speed} transmissions. Compared to communications with fixed infrastructures, UAVs face new challenges \rev{because of} their high altitude above the ground and great flexibility of movement in three-Dimensional (3D) space. Some critical issues include Line-of-Sight (LoS) dominant UAV-ground channels, the distinct communication Quality of Service (QoS) requirements for UAV control messages versus payload data, the stringent constraints imposed by the Size, Weight, and Power (SWAP) limitations of UAVs~\cite{zeng2019accessing}, to name a few.

\begin{figure}[!htb]
    \centering
    \includegraphics[width=\linewidth]{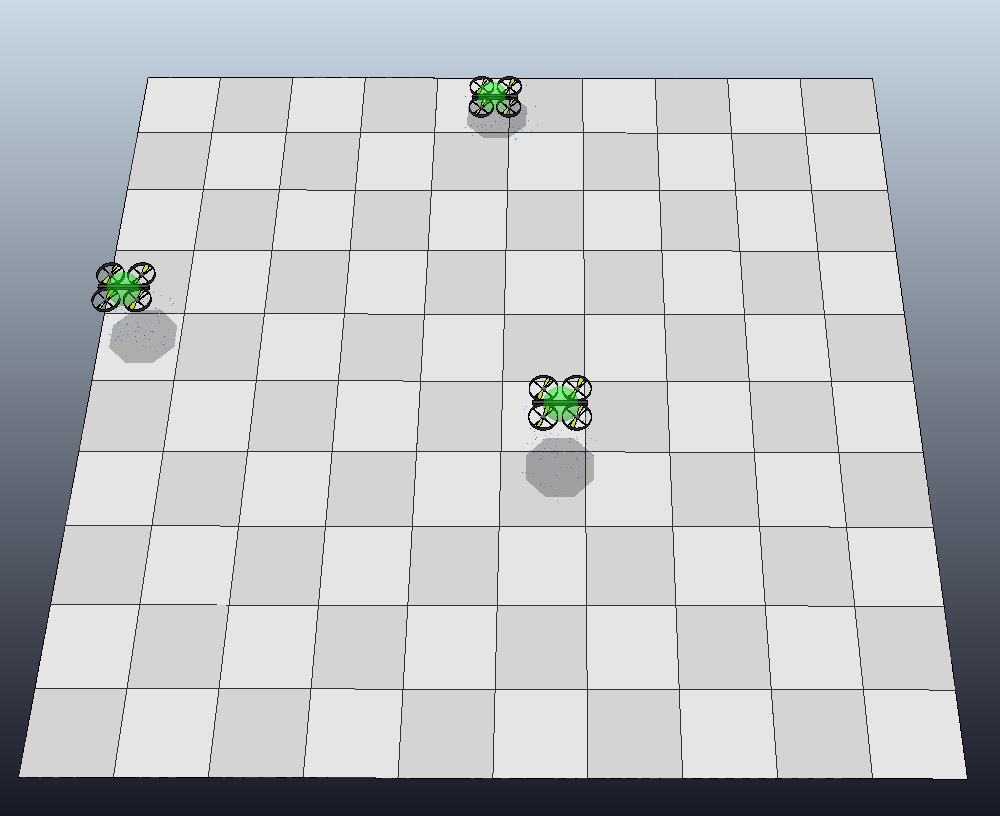}
    \caption{UAVs on CoppeliaSim.}
    \label{fig:coppelia}
\end{figure}

\ITUpar The application of UAV has played a significant role in civil applications and is rapidly expanding~\cite{alliance2020ran}, including agricultural plant protection, power inspection, police enforcement, geological exploration, and environmental monitoring. In this case study, the near-RT RIC \rev{makes } a radio resource allocation for on-demand coverage of UAVs, \rev{that takes account of} the radio channel conditions, flight path information, and other information \rev{about applications}.

\begin{figure}[!b]
    \centering
    \includegraphics[width=\linewidth]{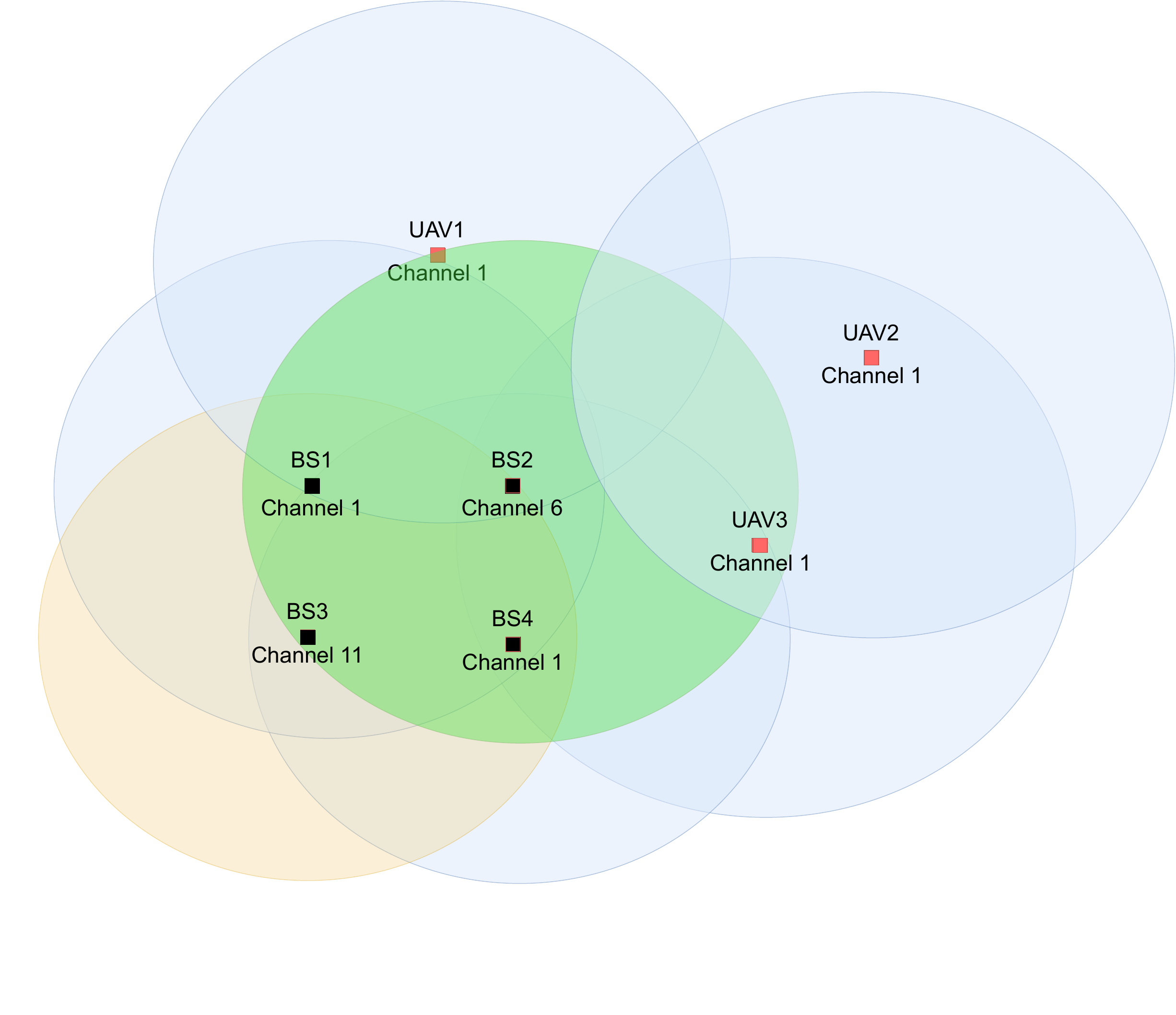}
    \caption{Network topology.}
    \label{fig:scenario1_general}
\end{figure}

\ITUpar For this second use case, we \rev{ran} Mininet-RAN with the robotics simulator CoppeliaSim\footnote{\url{https://www.coppeliarobotics.com/}} in a UAV flight path scenario \rev{which} is useful for setting the UAV paths (see Fig.~\ref{fig:coppelia}). The network topology consists of four BSs on land and three UAVs \rev{in} air. Each BS on land has its own pre-defined and static channel. UAVs also act as \rev{the} BS but with \rev{a} capacity to dynamically adjust their channel by receiving near-RT RIC control commands,  avoiding channel overlapping among them and BSs on land. 
Fig.~\ref{fig:scenario1_general} \rev{illustrates} the simulation deployment scenario, which comprises a Mininet-RAN topology with four BSs transmitting signals on channels 1, 6, and 11, and three UAVs on channel~1.

\begin{figure}[!htb]
\centering
    \includegraphics[width=\linewidth]{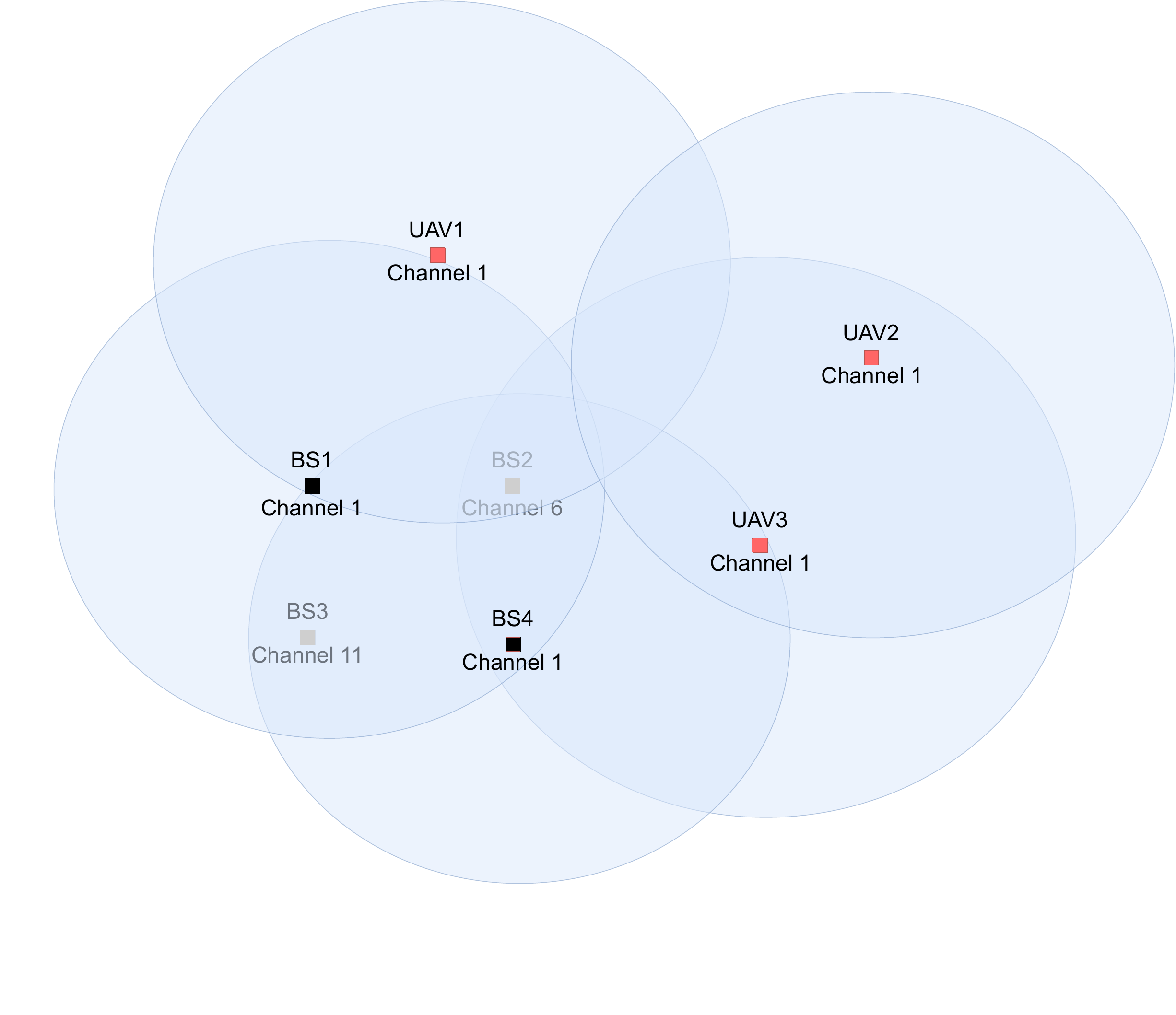}        
\caption{Signal overlapping illustration.}
\label{fig:overlapsch1}
\end{figure}

\ITUpar During the experiment it is expected to observe \rev{considerable} interference between \texttt{BS1} and \texttt{BS4} as well as among UAVs because they are transmitting signals on the same channel, as illustrated in Fig.~\ref{fig:overlapsch1}. As a result of the occupation of the same channel by multiple BSs on land and UAVs in the air, the communication using this channel will be impaired \rev{owing} to the high level of interference.

\ITUpar Our ML-powered xApps mitigate this problem by dynamically defining the UAV channels. We trained a model that can select the highest channel number distance \rev{between} BSs and UAVs, dynamically allocating UAVs' channels while they fly around. It \rev{should be noted} that this use case is not aiming to prove the ML algorithm efficiency but to demonstrate Mininet-RAN and near-RT RIC interaction capabilities. 

\ITUpar From now on, we \rev{will concentrate on} the second set of results, \rev{and discuss} the ML methods applied to the second use case, followed by the behavior of the monitored UAVs \rev{by means of} the implemented dashboard.

\subsubsection{Machine learning of use case 2}

\ITUpar The dataset used to train the ML for this use case has approximately 48000 samples, each consisting of the connection between one of three UAVs (\texttt{UAV1}, \texttt{UAV2} and \texttt{UAV3}) and one of four different sensors (\texttt{S1}, \texttt{S2}, \texttt{S3} and \texttt{S4}) attached to the BSs on land (\texttt{BS1}, \texttt{BS2}, \texttt{BS3} and \texttt{BS4}). \rev{These} sensors could measure the RSSI from UAVs and \rev{neighboring} BSs.

\ITUpar \rev{Since} the primary goal of the ML for this use case is to select the best channel for each UAV, it is \rev{necessary} to \rev{measure} the distance between the channel number used by the BS on land and the UAVs. The channel number \rev{ranges} from one to twelve, and the higher the channel number distance between them, the \rev{lower} the interference is, \rev{hence}, the better the UAVs’ connection. All of this information is \rev{arranged} in the dataset for each instant of time.

\ITUpar The dataset is \rev{divided} into 12 columns as follows: \texttt{Timestamp}: responsible for indicating the instant of time that the data is collected; \texttt{Tx\_BS\_S}: refers to the BSs; \texttt{Rx\_BS\_S}: represents the sensor at BSs; \texttt{Channel\_BS\_S}: indicates the channel used by that BS; \texttt{Tx\_DR\_S}: refers to the UAV under analysis; \texttt{Rx\_DR\_S}: represents the UAV channel; \texttt{Power\_DR\_S}: \rev{this} is the RSSI sensed by \rev{the} BS sensor from that UAV; \texttt{Old\_channel\_DR\_S}: indicates the current UAVs’ channel; \texttt{Old\_channel\_DR\_1}, \texttt{Old\_channel\_DR\_2} and \texttt{Old\_channel\_DR\_3}: represent, respectively, the channel where \texttt{DR1}, \texttt{DR2} and \texttt{DR3} are currently transmitting except for the UAV under analysis;  \texttt{New\_channel\_DR\_S}: represents the output of each sample, which refers to the best channel for the UAV under analysis. This output is encoded using the One hot encoding technique.

\begin{figure}[!tb]
    \centering
    \includegraphics[width=\linewidth]{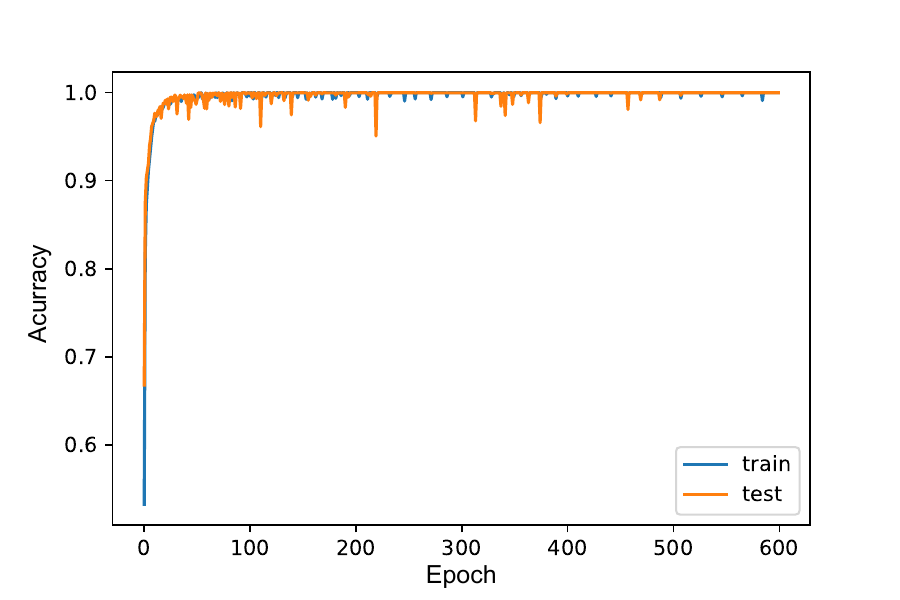}
    \caption{Learning curve.}
    \label{fig:learning_curve_use_case_2}
\end{figure}

\ITUpar As in \rev{use case} 1, we also \rev{used} an MLP with the following configuration: (i) four inputs, values of \texttt{Channel\_BS\_S}, \texttt{Old\_channel\_DR\_1}, \texttt{Old\_channel\_DR\_2} and \texttt{Old\_channel\_DR\_3}; and (ii) seven outputs: the probability of changing the UAVs’ channel for \texttt{Channel\_1}, \texttt{Channel\_2}, \texttt{Channel\_3}, \texttt{Channel\_5}, \texttt{Channel\_6}, \texttt{Channel\_11} and \texttt{Channel\_12}.

\ITUpar The preprocessed dataset is split for training and \rev{the} testing sets \rev{contain} 70\% and 30\% of the data, respectively. Fig.~\ref{fig:learning_curve_use_case_2} depicts the learning curve of the produced model, which refers to the performance \rev{of} the \rev{training and testing} data. As we \rev{achieved a}  similar \rev{degree of} accuracy for the \rev{training and testing} sets, the ML model has learned \rev{how to adapt to} the training data and is likely to perform well \rev{with} unseen data.

\begin{figure}[!htb]
\centering
\begin{subfigure}{0.49\linewidth}
    \includegraphics[width=\textwidth]{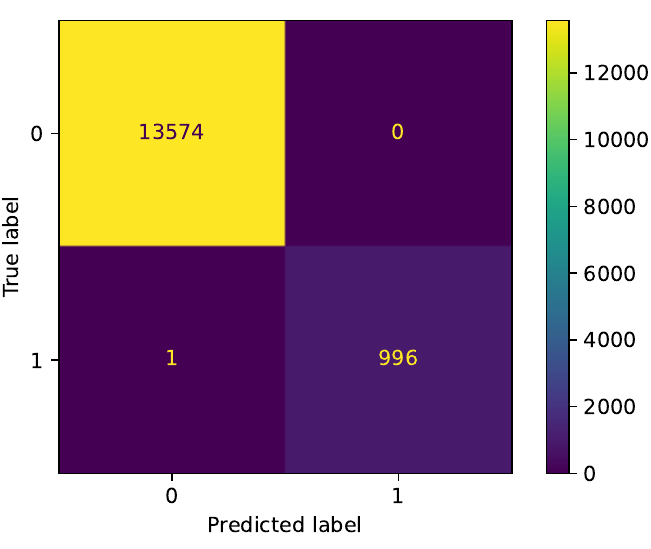}
    \caption{Channel 1.}
    \label{fig:Confusion_Matrix_Channel_1}
\end{subfigure}
\hfill
\begin{subfigure}{0.49\linewidth}
    \includegraphics[width=\textwidth]{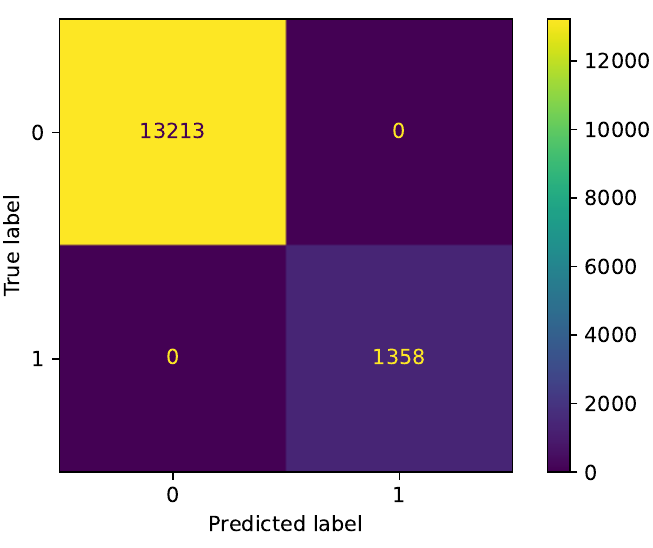}
    \caption{Channel 2.}
    \label{fig:Confusion_Matrix_Channel_2}
\end{subfigure}
\hfill
\begin{subfigure}{0.49\linewidth}
    \includegraphics[width=\textwidth]{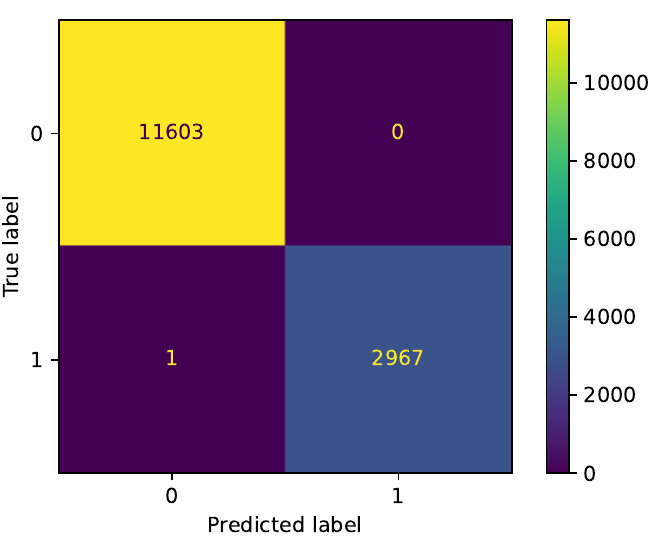}
    \caption{Channel 3.}
    \label{fig:Confusion_Matrix_Channel_3}
\end{subfigure}
\hfill
\begin{subfigure}{0.49\linewidth}
    \includegraphics[width=\textwidth]{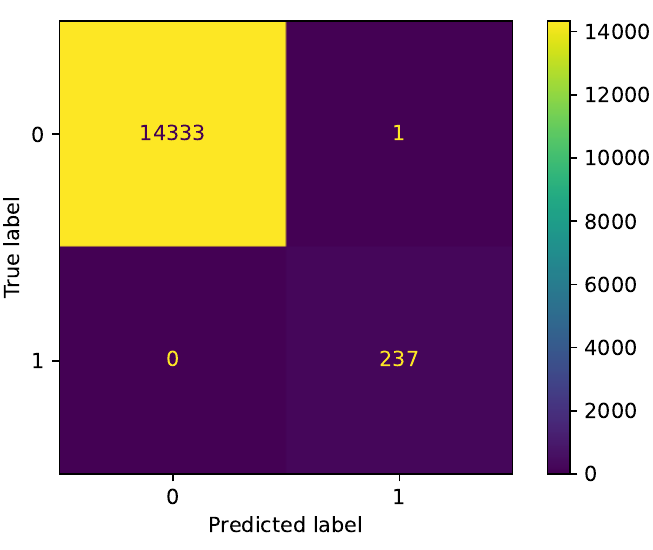}
    \caption{Channel 5.}
    \label{fig:Confusion_Matrix_Channel_5}
\end{subfigure}
\hfill
\begin{subfigure}{0.49\linewidth}
    \includegraphics[width=\textwidth]{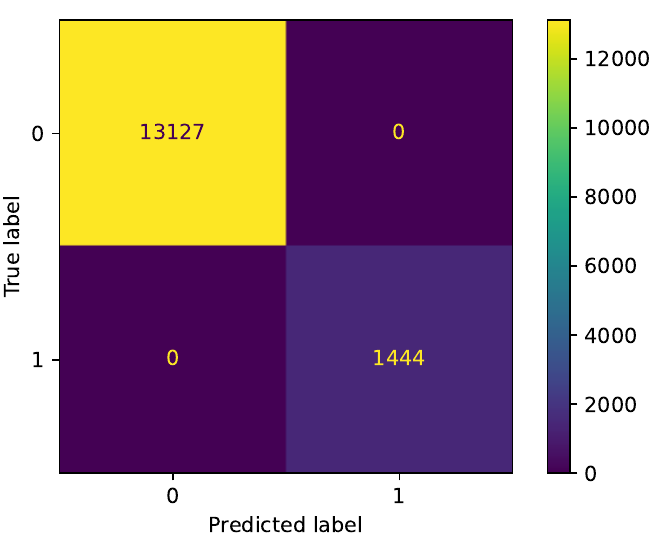}
    \caption{Channel 6.}
    \label{fig:Confusion_Matrix_Channel_6}
\end{subfigure}
\hfill
\begin{subfigure}{0.49\linewidth}
    \includegraphics[width=\textwidth]{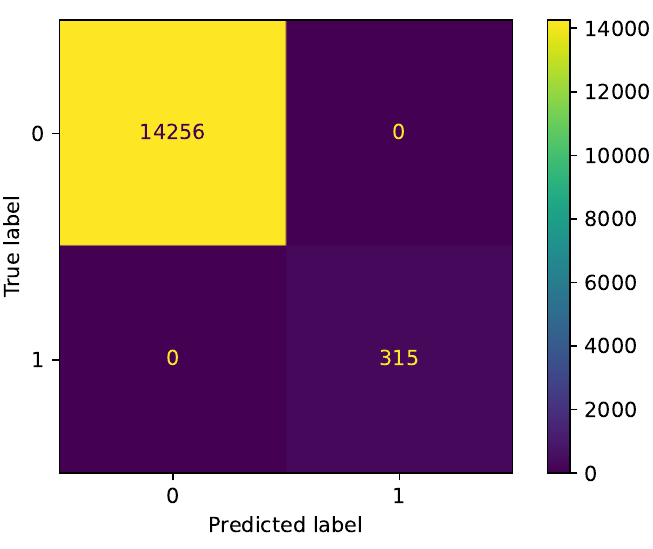}
    \caption{Channel 11.}
    \label{fig:Confusion_Matrix_Channel_11}
\end{subfigure}
\hfill
\begin{subfigure}{0.49\linewidth}
    \includegraphics[width=\textwidth]{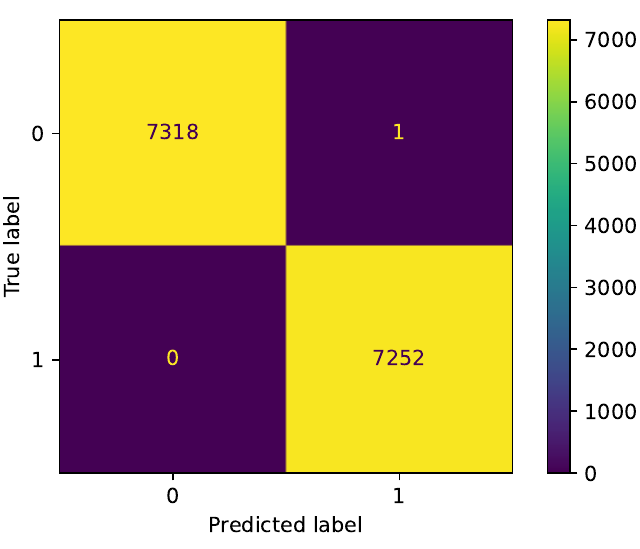}
    \caption{Channel 12.}
    \label{fig:Confusion_Matrix_Channel_12}
\end{subfigure}
        
\caption{Confusion matrices.}
\label{fig:confusion_matrix_use_case_2}
\end{figure}

\ITUpar The confusion matrices of the trained MLP \rev{are displayed}  in Fig.~\ref{fig:confusion_matrix_use_case_2} for \texttt{Channel\_1}, \texttt{Channel\_2}, \texttt{Channel\_3}, \texttt{Channel\_5}, \texttt{Channel\_6}, \texttt{Channel\_11} and \texttt{Channel\_12}. According to the results, ML \rev{made} four mistakes, indicating \texttt{Channel\_5} and \texttt{Channel\_12} when they were not the best option and \rev{failed} to indicate \texttt{Channel\_1} and \texttt{Channel\_3} when they were the best choice. Figures~\ref{fig:Confusion_Matrix_Channel_5}, \ref{fig:Confusion_Matrix_Channel_12},
\ref{fig:Confusion_Matrix_Channel_1} and \ref{fig:Confusion_Matrix_Channel_3} depict those errors, respectively.

\subsubsection{\rev{Results} of use case 2}

\ITUpar The dashboard \rev{designed} for this use case \rev{shows} the actual and the last measured power levels, as well as the channel of each BS and UAV in the emulated scenario.

\ITUpar As the UAVs move along their flight path \rev{which was} previously defined in the CoppeliaSim, the near-RT RIC keeps measuring KPIs to \rev{supply} ML-powered xApps. Here, the algorithm can recommend a better \rev{distance of the} channel number \rev{for} a UAV, triggering an action to change its channel by the near-RT RIC. As power levels and channel numbers change, the table of the implemented dashboard updates, keeping track of a brief record of the UAV channels and powers. We also could check the BSs power to \rev{confirm} the interference level \rev{changes because of} a better channel selection by the UAV. Table~\ref{tab:table-dashboard-us2} \rev{displays} this information.

\section{Limitations of the Study and Suggestion for Future Work}
\label{sec_limitations}

\ITUpar Mininet-RAN inherits all the limitations of the Mininet and Mininet-WiFi lightweight virtualization architecture, certainly adding a couple more related to our current approach. It is notorious that Mininet-RAN does not support 5G-related technologies, such as gNodeBs and the typical signaling signals for this network element. The E2 signaling, defined by O-RAN to communicate Mininet-RAN as a \rev{suitable} O-RAN E2 node, is on our roadmap and will be added to Mininet-RAN soon.

\begin{table}[!t]
\caption{Near-RT RIC dashboard monitoring information.}
\label{tab:table-dashboard-us2}
\resizebox{\columnwidth}{!}{%
\begin{tabular}{|c|c|c|c|c|c|}
\hline
\textbf{TX} & \textbf{RX} & \textbf{Old Power} & \textbf{Old Channel} & \textbf{New Power} & \textbf{New Channel} \\ \hline
BS1         & S1          & -80                & 1                    & -75                & 1                    \\ \hline
UAV1        & S1          & -50                & 7                    & -30                & 12                   \\ \hline
BS2         & S2          & -40                & 7                    & -40                & 7                    \\ \hline
UAV2        & S2          & -90                & 10                   & -90                & 10                   \\ \hline
UAV3        & S2          & -80                & 8                    & -80                & 8                    \\ \hline
\end{tabular}%
}
\end{table}

\ITUpar The future of 5G in Mininet-RAN will certainly \rev{involve  going} through the wwan\_hwsim Linux Kernel module as well as initiatives such as linux-wpan\footnote{https://linux-wpan.org/documentation.html} where some updates are already being made in the Kernel of the Linux operating system, \rev{although they are} not available in the Kernel main track yet.

\section{Conclusion}
\label{sec_conclusions}

\ITUpar This work presents Mininet-RAN as an emulation platform in support of low-cost 5G and O-RAN open-source experimentation facilities. We evaluated Mininet-RAN \rev{from the standpoint of} two O-RAN Alliance-defined use cases involving V2X and UAV. \rev{In particular}, we explored \rev{the question of} context-based dynamic handover management scenarios for V2X and flight path-based dynamic UAV resource allocation. Although our focus was not on exploring the efficiency of machine learning algorithms nor \rev{on proposing} new efficient techniques that can \rev{provide} new results to the scenarios explored in the use cases, the Mininet-RAN proved it was a low-cost and viable alternative for experimenting with new O-RAN-compliant algorithms and techniques. Unlike the related work, we found in the literature that Mininet-RAN \rev{is} the only solution that \rev{achieves} a complete PHY abstraction, \rev{and allows an} E2E evaluation with O-RAN interfacing.

\section*{Acknowledgment}
\ITUpar \rev{This research was partially funded by Lenovo, as part of its R\&D investment under Brazilian Informatics Law, and by the \textit{Coordenacao de Aperfeicoamento de Pessoal de Nivel Superior - Brasil} (CAPES) - Finance Code 001.}

\printbibliography 

\ITUpar

\begin{wrapfigure}{l}{25mm} 
 \includegraphics[width=1in,height=1.25in,clip,keepaspectratio]{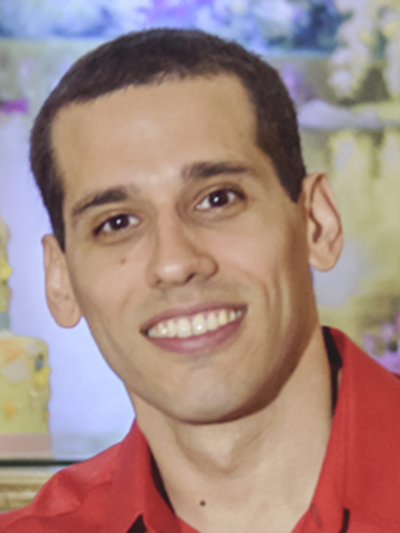}
  \end{wrapfigure}\par
  \textbf{Ramon dos Reis Fontes} is Professor in Computer Science at Federal University of Rio Grande do Norte (UFRN), Natal/Brazil. Before his current appointment he received a M.Sc degree from University of Salvador (UNIFACS), Salvador, Brazil, in 2014, and Ph.D. degree in Electrical and Computer Engineering from the State University of Campinas, SP, Brazil, in 2018. During his Ph.D., he developed a Wireless Network Emulator that has been widely used by researchers worldwide, and was a visiting PhD student at the DIANA team, in Sophia Antipolis/France in 2016. From Jul/2011 to Jun/2011, he was a professor at the Federal Institute of Science and Technology of Bahia (IFBA), where he taught subjects related to computer networks including network operating system, routers, application servers, information security, and programming. His main research interest are in the field of software defined wireless networks (SDWN), vehicular networking, future internet architecture and cyber security.\par

\begin{wrapfigure}{l}{25mm} 
 \includegraphics[width=1in,height=1.25in,clip,keepaspectratio]{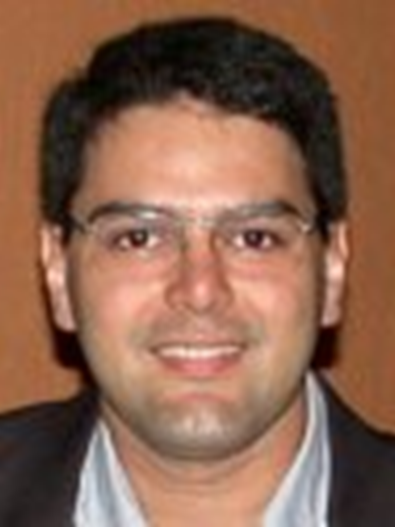}
  \end{wrapfigure}\par
  \textbf{Allan de Medeiros Martins} received the B.Sc. degree in electrical engineering, the M.E.Sc. degree in sciences, and the Ph.D. degree in electrical engineering from the Federal University of Rio Grande do Norte, Brazil. He is associated with the Department of Electrical Engineering, Federal University of Rio Grande do Norte. He does research in data mining, computing in mathematics, natural science, engineering and medicine and artificial neural networks.\par

\begin{wrapfigure}{l}{25mm} 
 \includegraphics[width=1in,height=1.25in,clip,keepaspectratio]{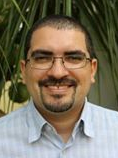}
  \end{wrapfigure}\par
  \textbf{Vicente Angelo de Sousa Junior} received his B.S., M.S and Ph.D. degrees in Electrical Engineering from the Federal University of Ceará (UFC), Fortaleza, CE, Brazil, in 2001, 2002 and 2009, respectively.  Between 2001 and 2006, he developed solutions to UMTS/WLAN interworking for   UFC   and   Ericsson of   Brazil. Between 2006 and 2010, he contributed to WIMAX standardization and Nokia’s product as a researcher at the Institute of Technological Development (INdT). Dr.  Sousa is now a professor and the head of the GppCom Research Group at UFRN. He is contributing to 5G open RAN projects supported by Lenovo and to NIR measurements and evaluation projects supported by ANATEL Agency.\par

\begin{wrapfigure}{l}{25mm} 
 \includegraphics[width=1in,height=1.25in,clip,keepaspectratio]{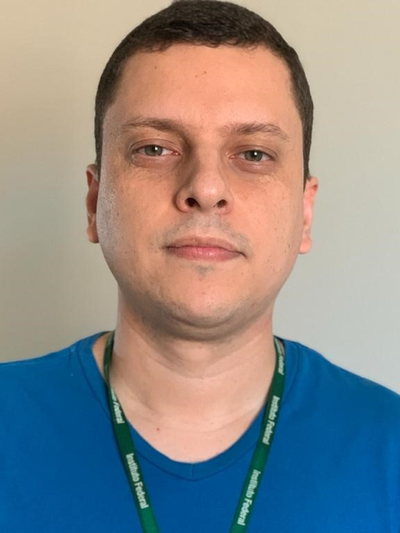}
  \end{wrapfigure}\par
  \textbf{Kaio Henrique Fonseca Dantas} received his A.S. in Computer Networks from the Federal Institute of Education, Science, and Technology of Rio Grande do Norte (IFRN) in 2013 and an MBA degree in Information Technology Management from Anhanguera University (UNIDERP) in 2019. Currently, He serves as a professor of computer networks at IFRN - Campus Pau dos Ferros. Additionally, he holds the position of Coordinator of IT Infrastructure at the Center for Data Analysis and Computational Intelligence (NADIC) from IFRN.\par
  
\begin{wrapfigure}{l}{25mm} 
 \includegraphics[width=1in,height=1.25in,clip,keepaspectratio]{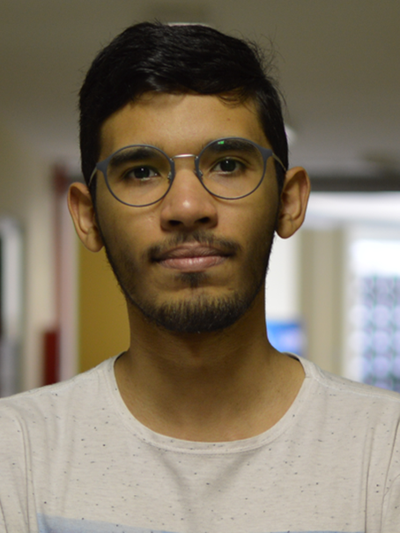}
  \end{wrapfigure}\par
  \textbf{Lucas Ismael Campos Medeiros}  received his B.S. degree in Telecommunications Engineering from the Federal University of Rio Grande do Norte (UFRN), Natal, RN, Brazil, in 2021. BSc. Medeiros is now pursuing a master's degree and Researching with GppCom Research Group at the Federal University of Rio Grande do Norte (UFRN), Brazil. He is contributing to 5G open RAN projects supported by Lenovo and to NIR measurements and evaluation projects supported by ANATEL Agency. \par
  
\begin{wrapfigure}{l}{25mm} 
 \includegraphics[width=1in,height=1.25in,clip,keepaspectratio]{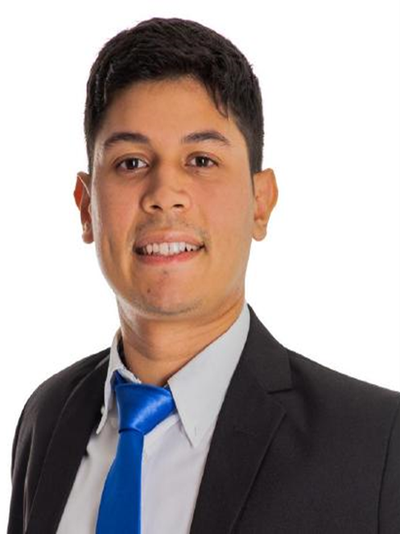}
  \end{wrapfigure}\par
  \textbf{Pedro Victor Andrade Alves} received the B.S. degree in computer engineering from the Federal University of Rio Grande do Norte (UFRN), Natal, in 2021, where he is currently pursuing the M.Sc. degree in electrical and computer engineering and a Team Member of the Research Group on Embedded Systems and Reconfigurable Computing (RESRC). He is contributing to 5G open RAN projects supported by Lenovo. His main research topic is tactile internet. His research interests include artificial intelligence, embedded systems, and tactile internet.\par
  
\begin{wrapfigure}{l}{25mm} 
 \includegraphics[width=1in,height=1.25in,clip,keepaspectratio]{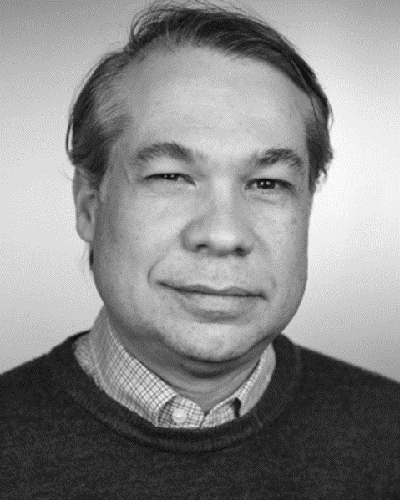}
  \end{wrapfigure}\par
  \textbf{Marcelo Augusto Costa Fernandes} Fernandes was born in Natal, Brazil. He received the B.S. and M.S. degrees in electrical engineering from the Federal University of Rio Grande do Norte (UFRN), Natal, in 1997 and 1999, respectively, and the Ph.D. degree in electrical engineering from the University of Campinas, Campinas, State of São Paulo, Brazil, in 2010. From 2015 to 2016, he worked as a Visiting Researcher with the Centre Telecommunication Research (CTR), King’s College London, London, U.K. From 2019 to 2021, he worked as a Visiting Scholar with the John A. Paulson School of Engineering and Applied Sciences, Harvard University, Cambridge, USA. He is currently an Associate Professor with the Department of Computer Engineering and Automation, Federal University of Rio Grande do Norte. He is the Leader of the Research Group on Embedded Systems and Reconfigurable Computing (RESRC), and a Coordinator of the Laboratory of Machine Learning and Intelligent System (LMLIS). His research interests include artificial intelligence, digital signal processing, embedded systems, reconfigurable hardware, and tactile internet.\par
  
\begin{wrapfigure}{l}{25mm} 
 \includegraphics[width=1in,height=1.25in,clip,keepaspectratio]{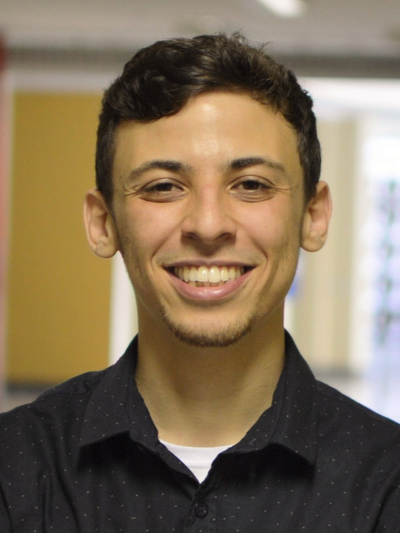}
  \end{wrapfigure}\par
  \textbf{Iago Diogenes do Rego} received his B.Sc. in Electrical Engineering and his M.Sc. in Electrical and Computing Engineering from the Federal University of Rio Grande do Norte (UFRN), Natal, in 2018 and 2020, respectively. He is currently pursuing a joint Ph.D. degree in Paris at the Engineering School of Information and Digital Technologies (EFREI Paris) and UFRN. He contributes to 5G Open RAN projects supported by Lenovo and his main research interests are in the field of mobile communications, 5G, resource allocation and machine learning.\par
  
\begin{wrapfigure}{l}{25mm} 
 \includegraphics[width=1in,height=1.25in,clip,keepaspectratio]{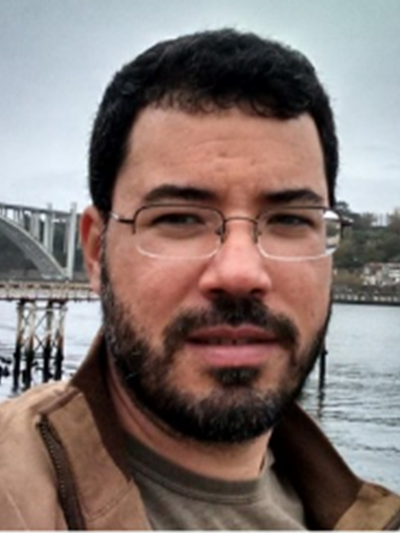}
  \end{wrapfigure}\par
  \textbf{Eduardo Aranha} received his B.S. (2000), M.S. (2002) and Ph.D. (2009) degree in Computer Science from the Federal University of Pernambuco, Brazil. He is currently an Associate Professor at the Informatics and Applied Mathematics Department (DIMAp), Federal University of Rio Grande do Norte (UFRN), Brazil. He worked as a Visiting Researcher at the School of Computer Science, University of Adelaide, Australia. His research interests include computer science education, software engineering, 5G and artificial intelligence applications.\par
  
\begin{wrapfigure}{l}{25mm} 
 \includegraphics[width=1in,height=1.25in,clip,keepaspectratio]{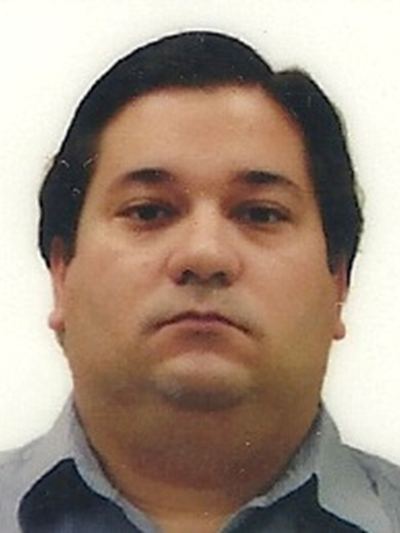}
  \end{wrapfigure}\par
  \textbf{Vinícius José Miranda Toscano de Brito Filho} received his B.Sc. (1999) degree in Computer Science from the Federal University of Pernambuco, Brazil. He is specialized in corporate network management by Santa Maria School (FSM) (2009), Pernambuco, Brazil, specialized in Information Technology by Federal University of Rio Grande do Norte (UFRN) (2021), Rio Grande do Norte, Brazil, M.B.A. in Business Management by Getúlio Vargas Foundation (FGV) (2017), Brazil. He contributes to 5G Open RAN projects supported by Lenovo and his main research interests are in the field of mobile communications, 5G, resource allocation and machine learning.\par
  
\begin{wrapfigure}{l}{25mm} 
 \includegraphics[width=1in,height=1.25in,clip,keepaspectratio]{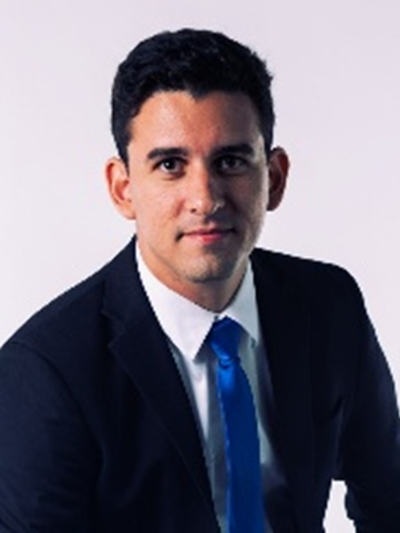}
  \end{wrapfigure}\par
  \textbf{Mateus Arnaud Santos de Sousa Goldbarg}  received a B.S. degree in Computer Engineering from the Federal University of Rio Grande do Norte (UFRN), Natal, RN, Brazil in 2021. As a member of the Laboratory of Machine Learning and Intelligent Instrumentation, since 2020 he researches techniques for compressing deep learning models for microservices and embedded systems. Currently, he is pursuing the M.Sc. degree in Electrical and Computer Engineering at UFRN and contributes to the 5G O-RAN project supported by Lenovo.\par
  
\begin{wrapfigure}{l}{25mm} 
 \includegraphics[width=1in,height=1.25in,clip,keepaspectratio]{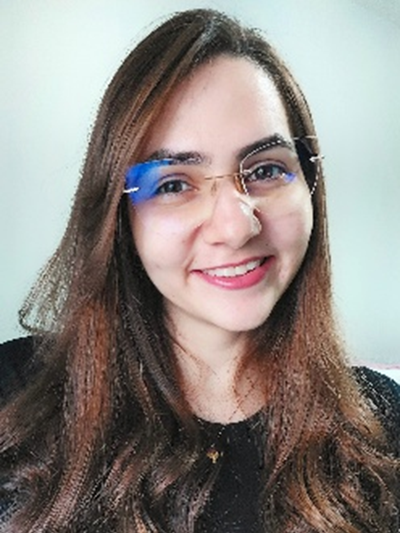}
  \end{wrapfigure}\par
  \textbf{Wysterlânya Kyury Pereira Barros} received her B.Sc. degree in Computer Engineering (2018) and M.Sc. degree in Electrical and Computing Engineering (2021) from the Federal University of Rio Grande do Norte (UFRN), Brazil. She is pursuing a Ph.D. in Electrical and Computing Engineering at UFRN and is a researcher of the Research Group on Embedded Systems and Reconfigurable Computing (RESRC). She contributes to 5G Open RAN projects supported by Lenovo, and her research interests include Artificial Intelligence, 5G Networks, Computer Vision, Embedded Systems, and Reconfigurable Hardware.\par
  
\begin{wrapfigure}{l}{25mm} 
 \includegraphics[width=1in,height=1.25in,clip,keepaspectratio]{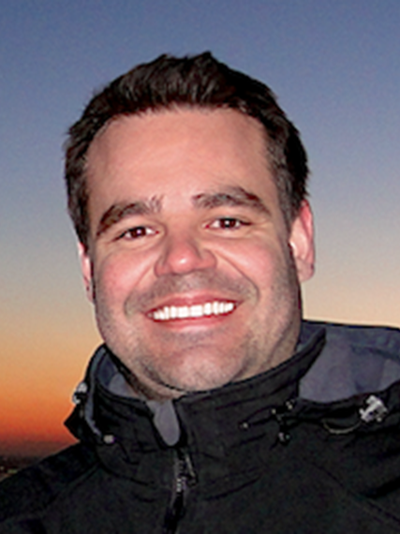}
  \end{wrapfigure}\par
  \textbf{Roger Immich} is a Professor at the Digital Metropolis Institute (IMD) of the Federal University of Rio Grande do Norte (UFRN). He earned his Ph.D. in Informatics Engineering from the University of Coimbra, Portugal (2017).  He was a visiting researcher at the University of California at Los Angeles, United States (UCLA) in 2016/2017, a postdoctoral researcher at the Institute of Computing of the University of Campinas (UNICAMP) in 2018/2019, and a Visiting Professor of University of Málaga, Spain, in 2021. Roger Immich is the author and co-author of over 75 articles published in highly reputable conferences and journals, both nationally and internationally. Additionally, he plays an active role in organizing conferences and workshops. His research interests encompass various areas, including Smart Cities, IoT, 5G, Quality of Experience, as well as Cloud and Fog computing.\par

\begin{wrapfigure}{l}{25mm} 
 \includegraphics[width=1in,height=1.25in,clip,keepaspectratio]{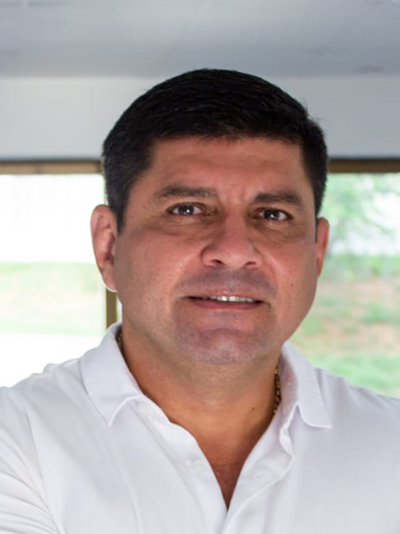}
  \end{wrapfigure}\par
  \textbf{Augusto Venancio Neto} Augusto Venâncio Neto (Senior Member, IEEE) is an Associate Professor at the Department of Informatics and Applied Mathematics of the Federal University of Rio Grande do Norte (UFRN), Natal/RN, Brazil; Productivity fellow at the National Council for Scientific and Technological Development (CNPq); and member of the Instituto de Telecomunicações, Portugal. He is the leader of the Research Group in Future Internet Service and Applications (REGINA), with background in cutting-edge technologies in computer networks and telecommunications, through which managing large research and developing teams, allowed him to form tens of PhDs and masters’ researchers, as well as for authoring/coauthoring more than 180 papers, including patents and several best-paper awards. He is involved in the organization of several international conferences and workshops, as well as serving as Guest Editor for special issues of peer-reviewed scholarly journals. His research interests fall in the fields of 5G/6G Networks, AI-supported networking, Mobile Computing, Smart Spaces, SDN, NFV and Cloud Computing.\par

\end{document}